\documentclass{aa}  

\usepackage{graphicx}
\usepackage{txfonts}
\usepackage{subfigure}
\usepackage[normalem]{ulem}

\begin{document} 

    \title{Constraining the polarisation flux density and angle of point sources by training a convolutional neural network}
    \titlerunning{PS polarisation using CNN}
    \authorrunning{J. M. Casas et al.}
   
    \author{J. M. Casas \inst{1,2},
    L. Bonavera \inst{1,2},
    J. Gonz{\'a}lez-Nuevo \inst{1,2},
    M. M. Cueli \inst{1,2},
    D. Crespo \inst{1,2}, 
    E. Goitia \inst{1},
    C. González-Gutiérrez \inst{2,3},
    J. D. Santos \inst{1,2},
    M. L. S{\'a}nchez \inst{1,2},
    F. J. de Cos \inst{2,4}
}

   \institute{$^1$Departamento de F{\'i}sica, Universidad de Oviedo, C. Federico Garc{\'i}a Lorca 18, 33007 Oviedo, Spain\\
    \email{casasjm@uniovi.es}\\
             $^2$Instituto Universitario de Ciencias y Tecnolog{\'i}as Espaciales de Asturias (ICTEA), C. Independencia 13, 33004 Oviedo, Spain\\
             $^3$Departamento de Inform{\'a}tica, Universidad de Oviedo, Edificio Departamental 1. Campus de Viesques s/n, E-33204, Gij{\'o}n, Spain\\
             $^4$Escuela de Ingeniería de Minas, Energía y Materiales Independencia 13, 33004 Oviedo, Spain
             }


 \abstract{}{}{}{}{} 
  \abstract
   {Constraining the polarisation properties of extragalactic point sources is a relevant task not only because they are one of the main contaminants for primordial cosmic microwave background B-mode detection if the tensor-to-scalar ratio is lower than r = 0.001, but also for a better understanding of the properties of radio-loud active galactic nuclei.}
   {We develop and train a machine learning model based on a convolutional neural network to learn how to estimate the polarisation flux density and angle of point sources embedded in cosmic microwave background images knowing only their positions.}
   {To train the neural network, we used realistic simulations of patches of 32$\times$32 pixels in area at the 217 GHz \textit{Planck} channel with injected point sources at their centres. The patches also contain a realistic background composed of the cosmic microwave background signal, the Galactic thermal dust, and instrumental noise. We split our analysis into three parts: firstly, we studied the comparison between true and estimated polarisation flux densities for $P$, $Q,$ and $U$ simulations. Secondly, we analysed the comparison between true and estimated polarisation angles. Finally, we studied the performance of our model with the 217 GHz \textit{Planck} map and compared our results against the detected sources of the Second \textit{Planck} Catalogue of Compact Sources (PCCS2).}
   {We find that our model can be used to reliably constrain the polarisation flux density of sources above the 80 mJy level. For this limit, we obtain relative errors of lower than 30$\%$ in most of the flux density levels. Training the same network with $Q$ and $U$ maps, the reliability limit is above $\pm$250 mJy when determining the polarisation angle of both $Q$ and $U$ sources. Above that cut, the network can constrain angles with a 1$\sigma$ uncertainty of $\pm29^{\circ}$ and $\pm32^{\circ}$ for $Q$ and $U$ sources, respectively. We test this neural network against real data from the 217 GHz \textit{Planck} channel, obtaining similar results to the PCCS2 for some sources; although we also find discrepancies in the 300-400 mJy flux density range with respect to the {\textit Planck} catalogue.}
   {Based on these results, our model appears to be a promising tool for estimating the polarisation flux densities and angles of point sources above 80 mJy in any catalogue with very small computational time requirements.}


   \maketitle
%

\section{Introduction}
\label{sec:introduction}

The impact of polarised extragalactic radio sources (ERSs), mainly blazars \citep{deZOT09, TUC11}, is not only important for research into active galactic nuclei but also because they are a contaminant for future cosmic microwave background (CMB) experiments as their properties are not
yet well constrained. 

ERSs are usually detected in total intensity signal $I$ and in both $Q$ and $U$ maps of the microwave sky. However, it is better to work with the total polarisation of a source, $P = \sqrt{Q^{2} + U^{2}}$ , not only because $P$ is definite positively, but also because it is physically related to the real processes occurring along the path of photons from the ERS to Earth \citep{HER12}. However, although blind and non-blind signal-processing techniques work by constraining the P polarisation, it is more common to present results of the total polarisation fraction, $\Pi = P/S$, where $S$ is the source flux density in total intensity. 

At present times, only a few sources are found to have a maximal polarisation fraction of  $\sim$10\%, a fact predicted and explained by \citet{TUC12}. Indeed, only 243 sources above the 99.99\% C.L. are detected in the Second Planck Catalogue of Compact Sources \citep[PCCS2,][]{PCCS2}, which corresponds to only 2.715\% of the sources detected in total intensity. 

The polarisation fraction value for HFI \textit{Planck} channels has been studied by applying the stacking technique to the mission data in \citet{Bon17a} for radio sources (mean value of 3.08\% at 217 GHz) and in \citet{Bon17b} for dusty galaxies (mean value of 3.10\% and 3.65\% at 217 and 353 GHz, respectively). More recently, \citet{GAL18} found a mean value of 2.91$\pm$0.42\% for flat-spectrum sources at 100 GHz and \citet{TRO18} found a median value of 2.83\% for radio sources and set upper limits (at 3.9\% and 2.2\%) for dusty galaxies (at 217 and 353 GHz, respectively) by exploiting the intensity distribution analysis. Furthermore, for the Atacama Cosmology Telescope Polarisation survey (ACTPol, \citealt{ACT19}), the authors conclude that the polarisation of ERS could be defined by a Gaussian distribution instead of a log-normal one, presenting a mean fractional polarisation of 2.9$\pm$0.5\% at 148 GHz, while for the South Pole Telescope Polarisation survey (SPTpol, \citealt{SPT19}), they obtain a weighted mean fractional polarisation of 2.63$\pm$0.22\% at 150 GHz.

The other physical quantity used to describe the behaviour of polarised ERS is the polarisation angle (also known as position angle), which is usually defined as
\begin{equation}
\begin{split}
    \psi \hspace{1pt} = \hspace{1pt} tan^{-1} \hspace{1pt} \left(\frac{U}{Q}\right), 
\label{eq:polarization_angle}
\end{split}
\end{equation}
where $0 \leq \psi \leq 2\pi$. As for the polarisation fraction, it is believed to be frequency independent \citep{SAI84}. Furthermore, although for lobe-dominated quasars the alignment between the core polarisation and the radio axis might be $\Delta \phi_{rad} \sim 90^{\circ}$ \citep{SAI84}, no constraints have yet been made on the polarisation angle of core-dominated quasars \citep{PEA84}.

Another important goal of the scientific community is to accurately constrain the polarised properties of ERSs as they directly affect primordial B mode detection, that is, the discovery of CMB patterns related to gravitational waves produced by cosmological perturbations at inflation era \citep[][]{STA79,LIN82}. Furthermore, following the simulations by \citet{REM18}, unresolved polarised ERS could be one of the main contaminants of such detections: as also confirmed by \citet{PUG18}, if the tensor-to-scalar ratio is lower than r = 0.001, ERSs affect the CMB power spectrum at multipoles $l$ > 50.

Several methods have been developed in recent years for the detection and flux density estimation of the polarisation of ERSs: \citet{ARG09} developed Filtered Fusion (FF), which was applied by \citet{LOP09} to both WMAP and \textit{Planck} data. More recently, \citet{HER21} developed a Bayesian estimator (BFF), reaching reliable polarised estimations down to 0.01 Jy in realistic simulations of the QUIJOTE experiment \citep{Rub12}. Furthermore, \citet{DP21} proposed a method based on steerable wavelets, which is reliable down to 3.38 and 5.76 mJy for regions of faint galactic emission in simulations of the 30 and 155 GHz bands, respectively, of the PICO experiment \citep{Han19}.

In this work, we present a new methodology based on a convolutional neural network \citep[CNN,][]{LeC89} called the POint Source Polarisation Estimation Network (POSPEN) trained with realistic simulations of polarisation data from the \textit{Planck} mission. In Section 2, we present simulations for training and validating the network. Section 3 presents details of POSPEN. Section 4 discusses the results from simulations and real data and in Section 5 we outline our conclusions.


\section{Simulations}
\label{sec:simulations}

\begin{figure*}[ht]
\centering
\includegraphics[width=11.5cm, height=3.5cm]{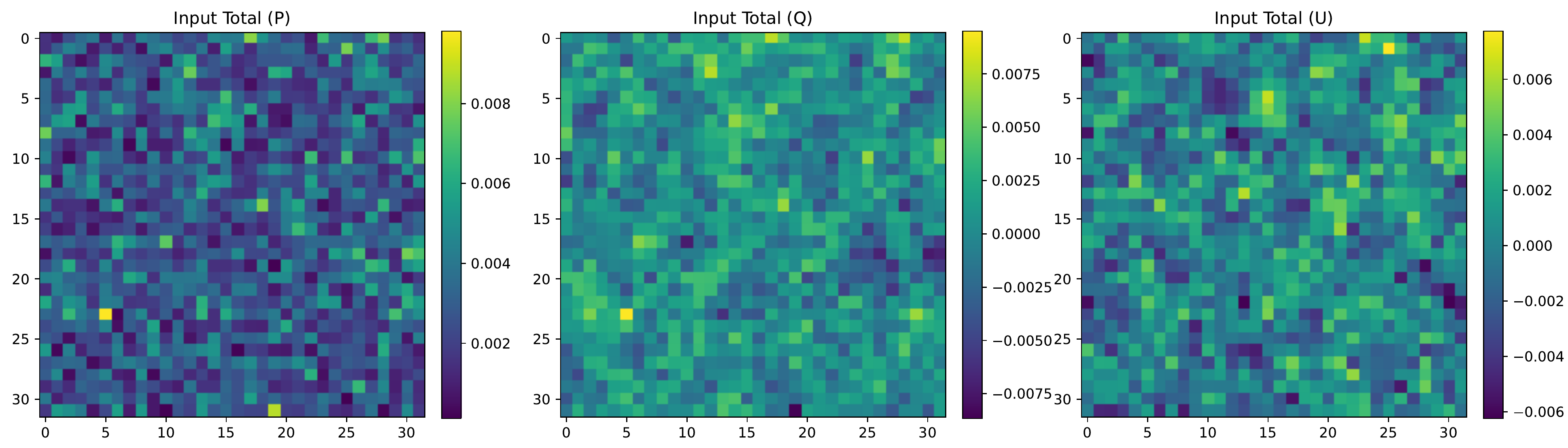}
\includegraphics[width=11.5cm, height=3.5cm]{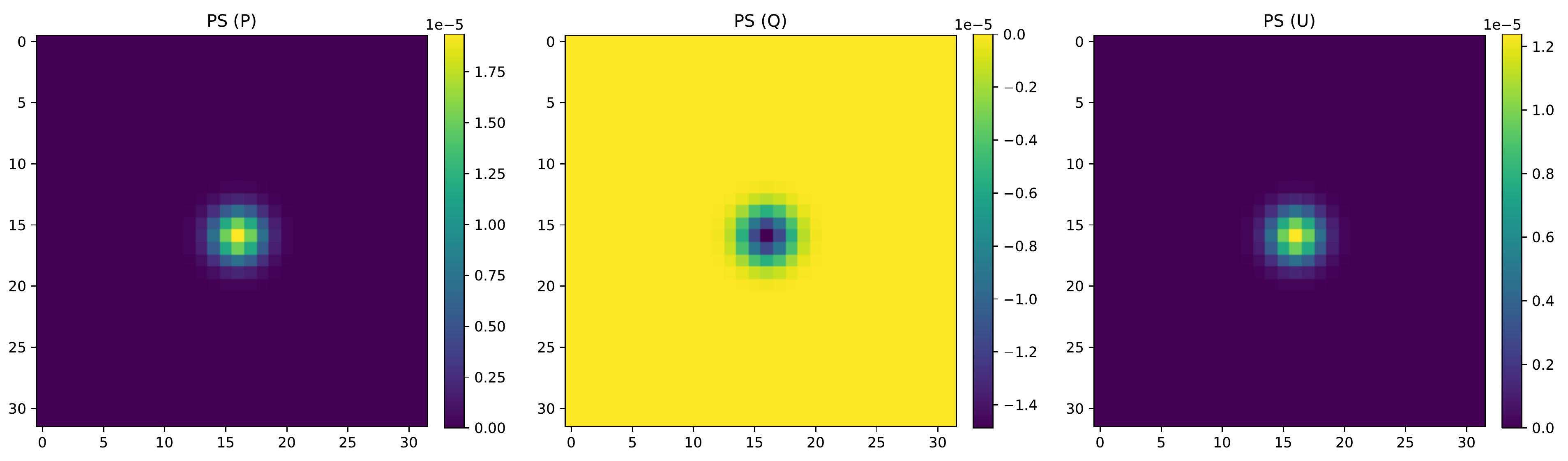}
\caption{Example of one simulation used to train or validate POSPEN for $P$, $Q,$ and $U$ polarisation maps (left, middle, and right columns, respectively). Top panels show the input patch formed by all the emissions, where the network learns how to extract the polarisation of the PS. Bottom figures show an example of the simulated PS, where we use the integrated intensity as a label for the training. The units are in Jy.}
\label{Fig:simulations}
\end{figure*}

For training and validating POSPEN, we use realistic simulations of the polarised sky at the 217 GHz \textit{Planck} channel from the PR3 release, which can be downloaded from the Planck Legacy Archive (PLA\footnote{http://pla.esac.esa.int/pla/\#home}) database. Our goal is to develop a methodology to constrain both the polarisation emission and angle for each point source (PS) in a given catalogue (i.e. their positions are known). In particular, we randomly divide the sky maps into patches of 32$\times$32 pixels in area and a pixel-side of 90 arcsec, which is reasonably close to the \textit{Planck} 1.72 arcmin \citep[corresponding to $N_{side}=2048$ in the \texttt{HEALPix} all-sky pixelisation schema;][] {GOR05}.

Each simulated patch has a central injected PS in order to mimic the future application of this methodology to real data where the PS position is known from its detection in total intensity (non-blind method). Its polarisation integrated value is kept to be used during the POSPEN training in the minimisation of the loss function. In addition, the patch contains realistic sky emission at the chosen frequency, a background consisting of the CMB signal, the Galactic thermal dust, and instrumental white noise at \textit{Planck} levels.

Although polarised dusty galaxies are also present at 217 GHz \citep{PCCS2}, for simplicity we consider only radio sources. These sources are simulated in total intensity following the C2Ex model by \citet{TUC11} and the software CORRSKY \citep{GN05}; they are then convolved with a Gaussian filter using the 4.90 arcmin full width at half maximum value of \textit{Planck} at this frequency. Their $P$ polarisation is estimated by assuming a log-normal distribution with the parameters set to $\mu = 0.7$ and $\sigma = 0.9$ \citep{Bon17a}. The polarisation angle ($\psi$) is assumed to be randomly oriented in any direction of the sphere. Then, following the convention by \citet{HAM96}, $P$ and $\psi$ are used to calculate both $Q$ and $U$ values as
\begin{equation}
\begin{split}
    Q \hspace{1pt} = \hspace{1pt} P \hspace{1pt} cos \hspace{1pt} (\psi),\\
    \vspace{1pt}
    U \hspace{1pt} = \hspace{1pt} P \hspace{1pt} sin \hspace{1pt} (\psi).
\label{eq:Q_U_expressions}
\end{split}
\end{equation}

The $Q$ and $U$ CMB maps are the ones from the \texttt{SEVEM} component-separation method \citep{MartinezGonzalez2003} and the Galactic thermal dust map is a simulation from the Planck Sky Model \citep{Del13} based on the FFP10 sky model\footnote{The acronym FFP stands for full focal plane, which is the parallel cross-instrument set of codes of all detectors at all frequencies used by the \textit{Planck} collaboration.}. The instrumental noise is added to the patch as a random white noise using the \textit{Planck} value of 1.75 $\mu K_{CMB}$ for this channel \citep{PLA_18_I}, after re-scaling this value using the pixel area and the simulated beam. 

In this work, we make two main groups of simulations: one in $P$ and one in $Q$ and $U$. First, we simulate 10000 patches to train POSPEN in $P$ which allow us to work with positive defined polarisation values for the simulated sources because it is unclear at this point whether or not the neural network can properly handle positive and negative data in the same dataset. Then, to take  the possibility of overfitting into account, we also simulate 1000 extra patches for testing the data during training. Moreover, we prepare a validation dataset of 1000 additional patches not used in the POSPEN training. To prepare the second group of simulations, we repeat the same procedure for the train, test, and validation datasets formed by $Q$ and $U$ patches. An example of the simulated maps is shown in Figure \ref{Fig:simulations}.


\section{Methodology}
\label{sec:methodology}

\begin{figure*}[ht]
\centering
\includegraphics[width=17cm,height=7.25cm]{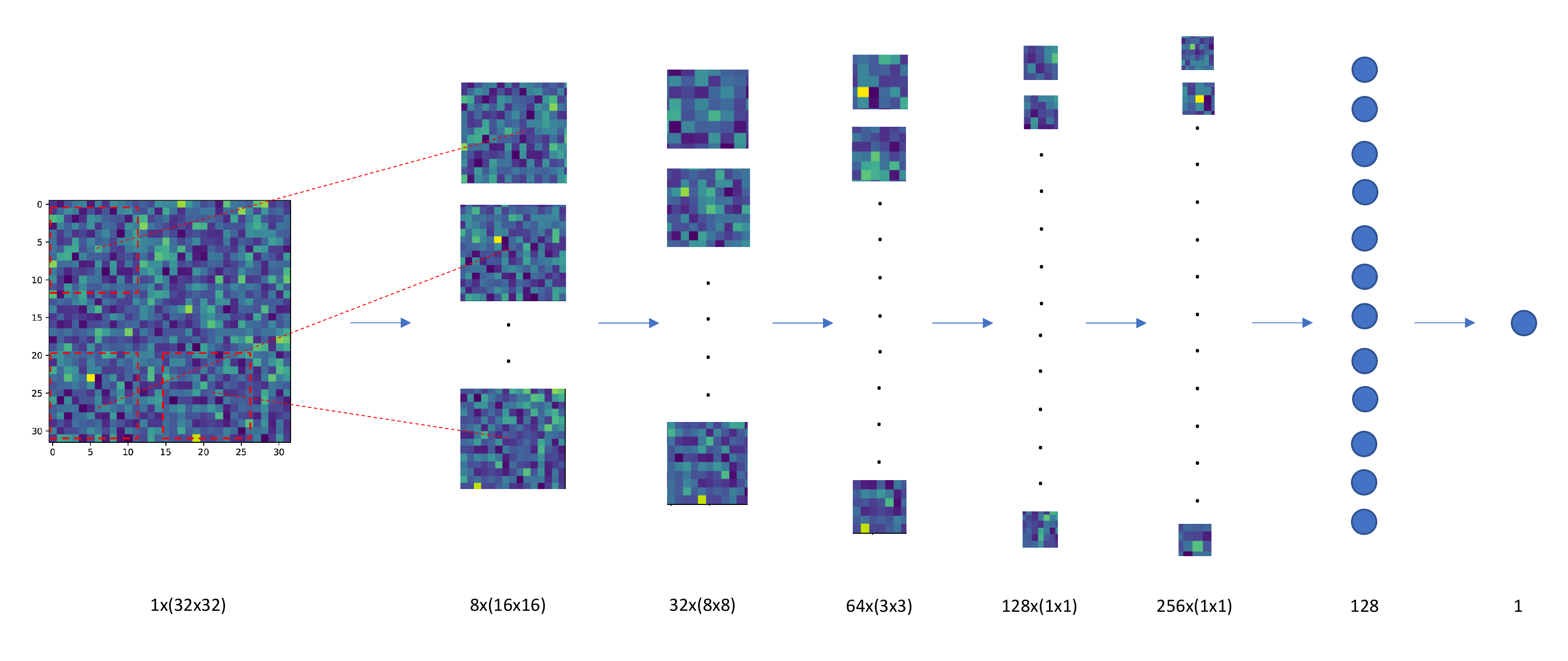}
\caption{Architecture of POSPEN. The information in the 217 GHz patch is convolved into 8 filters in the first convolutional block. After that, several convolutions are made through four more layers of 32, 64, 128, and 256 filters. The information is then processed through a layer of 128 neurons. Finally, one last neuron gives the output value.}
\label{Fig:architecture}
\end{figure*}

In recent years, as the quantity of available data has continuously increased, the impact of machine learning (ML) has also increased across a vast range of technologies. One of the most popular ML methods is the neural network. These are models based on neuroscience and are specially designed to learn non-linear behaviours in data \citep[see][and references therein]{GOO16}. Neural networks are formed by layers of weighted computational units called neurons. Their weights are updated upon each step of training: once the information flowing forward reaches the last layer, a loss function is minimised and a gradient is computed in order to adjust the weights by flowing the information  back and retrospectively updating each layer, one by one, back to the first layer. This process is called backpropagation \citep{Rum86} and the whole process of forward and backward flows of information through the architecture is called an epoch of training. This process is iterated in order to optimise the network and  several epochs can be required to train the network.

New methods based on neural networks and other ML methods have
recently been developed in the field of CMB research, with promising results being obtained in both total intensity and polarisation. For example, \citet{BON21} and \citet{CAS22} compared their fully convolutional neural networks to commonly used filters in \textit{Planck} for PS detection in single-frequency \citep{GN06} and multi-frequency \citep{Her09}  realistic total-intensity simulations, obtaining more reliable results, and also at frequencies not used for training the networks. 

In component separation, \citet{CAS22b} recently developed a fully convolutional neural network for extracting the CMB signal in total intensity in realistic \textit{Planck} simulations, reaching low residual values for multipoles up to 4000. These authors found that those types of architectures might be able to deal with both foreground and systematic non-gaussianities in future polarisation studies. A similar conclusion was derived by \citet{KRA21}, who proposed a model based on generative adversarial networks capable of creating non-Gaussian simulations of polarised thermal dust extended to 12 arcmin using low-resolution input data. Moreover, \citet{KRA19} also investigated a methodology based on a neural network for extending a CNN to the sphere, and \citet{PUG20} developed a model based on convolutional neural networks for inpainting Galactic foregrounds.

Here, we propose the use of CNNs for estimating the polarisation flux density and angle values of sources previously detected in total-intensity maps. These kinds of networks are composed of blocks: each block convolves the input information into a space formed by a determined number of feature maps. The information is then computed for each block by a group of weights called kernels. After several convolutions, one or more layers of several neurons take into account the relevant convolved information by running a non-linear activation function, and a final neuron connected to them provides an estimated value. At the end of this architecture, a loss function is minimised using the value given as the label and the gradient is computed using this loss information. With the gradient estimation, the weights of each convolutional block are updated on each layer of the architecture in the backpropagation \citep{Rum86}. This process \citep[explained in detail in ][]{GOO16} is iterated for several epochs.

In our case, the input data are a patch of the microwave sky read by the first convolutional block as a 32$\times$32 square matrix. The information is then convolved along five convolutional blocks, formed by 8, 32, 64, 128, and 256 kernels\footnote{The number of kernels and therefore their sizes have been selected in order to consider the final convolved information from a feature space equivalent to 1 pixel.}, with sizes of 2, 2, 3, 5, and 5 and strides\footnote{This is the number of pixel shifts over the input matrix, i.e. the network moves the filters n pixels at a time on each layer.} of 2, 2, 3, 5, and 5. After the convolutional blocks, there are two layers of 128 and 1 neurons, respectively, which convert the previous information into numerical values. In all layers, the selected activation function is the leaky rectified linear unit \citep[leaky ReLU,][]{NAI10}, which   not only allows the information   to pass when the neurons of each block are active but also applies a small gradient when they are not active. This last behaviour is controlled by an alpha parameter, which in our case is 0.2. The other selected hyperparameters are the AdaGrad optimiser, 500 epochs, a batch size of 16, and a learning rate of 0.05. These have been selected bqsed on a preliminary study before the final training of the network. Finally, in the last part of the architecture, we use the mean squared error loss function, defined as
\begin{equation}
    MSE \hspace{1pt} = \hspace{1pt} \frac{1}{2} \hspace{1pt} | \hspace{1pt} y \hspace{1pt} - \hspace{1pt} y^{\prime} \hspace{1pt} |\hspace{1pt}^{2},
\label{eq:mse}
\end{equation}
where $y$ is the value estimated by the network and $y^{\prime}$ is the true value we want to obtain. The architecture of POSPEN is shown in Figure \ref{Fig:architecture}. 

\section{Results}
\label{sec:results}

\subsection{Polarisation flux density estimation}
\label{sec:results_P}

In this section, we assess the results of our total polarisation flux density estimations, $P$, for the PS in the validation dataset. We compute the correlation between true and estimated polarisation, the polarisation relative error, and the source number counts in polarisation. In particular, the polarisation relative error is defined as
\begin{equation}
    PE \hspace{1pt} = \hspace{1pt} \frac{P_{estimated} \hspace{1pt} - \hspace{1pt} P_{true}}{P_{true}} \hspace{1pt} \times \hspace{1pt} 100 \hspace{1pt},
\label{eq:DP}
\end{equation}
where $P_{estimated}$ and $P_{true}$ are the estimated and true polarisation values, respectively.

\begin{figure*}[t]
\centering
\minipage{0.5\textwidth}
\includegraphics[width=\linewidth]{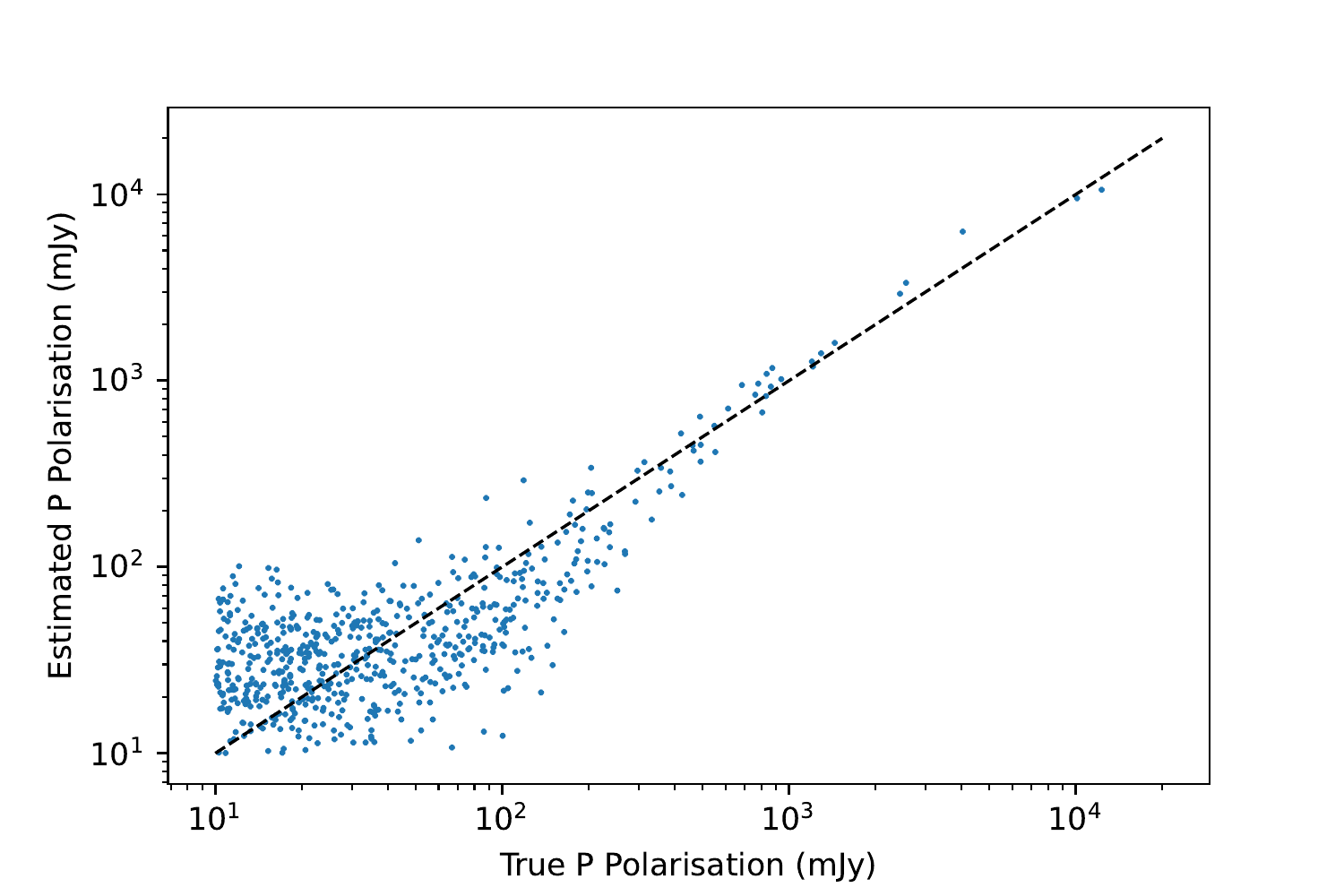}
\endminipage\hfill
\minipage{0.5\textwidth}%
  \includegraphics[width=\linewidth]{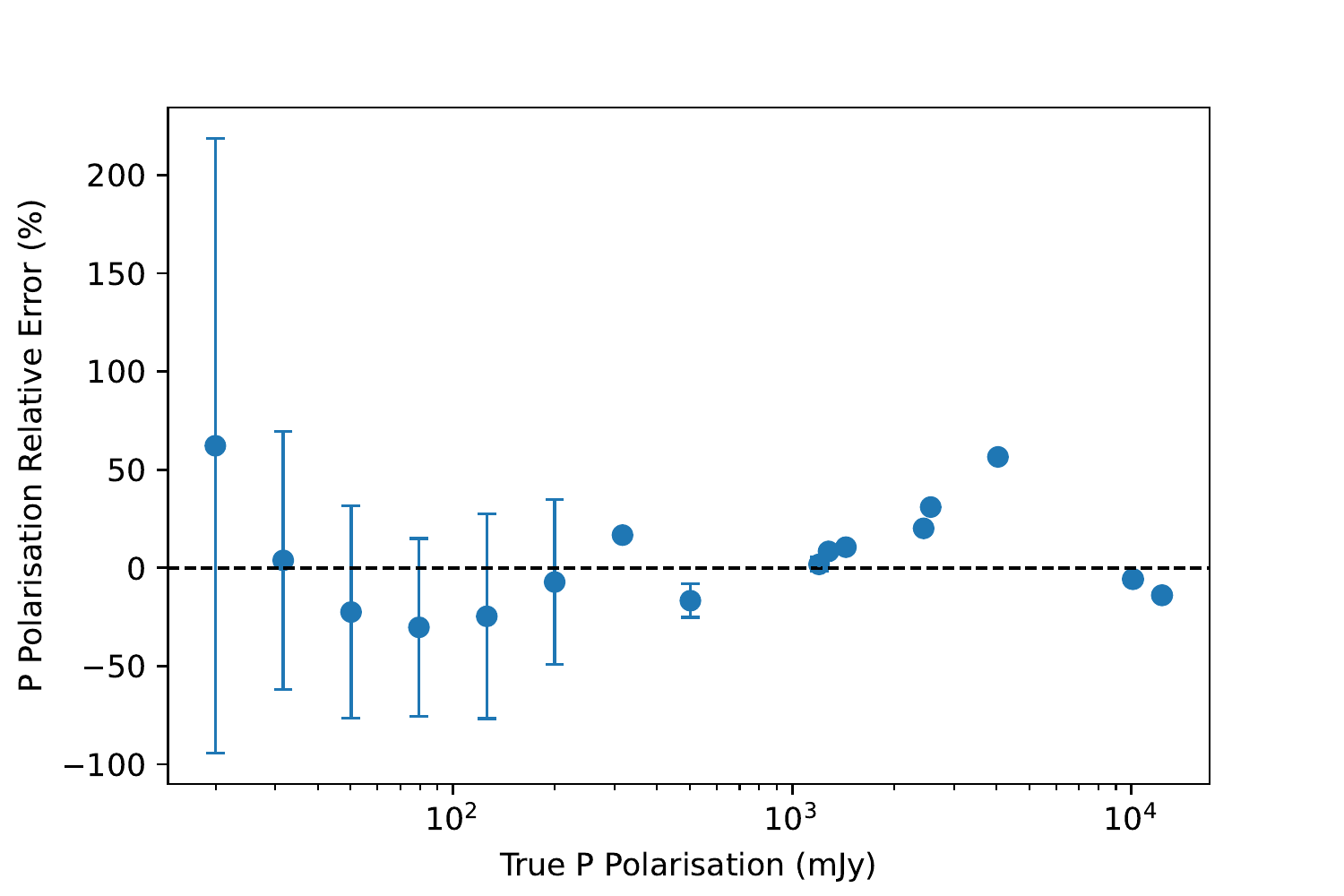}
\endminipage
\caption{Estimation of total polarisation flux density of the validation sources. Left panel: Correlation between true and estimated polarisation flux density. Right panel: Polarisation relative error against the input polarisation flux density. The error bars represent the uncertainty of each bin, which we considered to be equal to the standard deviation. In both cases, the dashed black lines show the ideal case.}
\label{Fig:correlation_P}
\end{figure*}

The left panel of Figure \ref{Fig:correlation_P} shows the correlation between true and estimated polarisation flux densities. The right panel plots the polarisation relative error against the true polarisation. In particular, we compute the mean relative error values for each bin (blue dots) and their uncertainty (standard deviation). In both panels, the ideal case is represented with dashed black lines.

Both figures show that the estimations by POSPEN for all simulated patches seem to be reliable above 80 mJy. More precisely, above that value, the CNN estimates the polarisation of the sources  relatively well, with an error of less than 30\% in most cases. Below 80 mJy, the results are affected by the Eddington bias: the estimated values are systematically overestimated, showing relative errors above 100$\%$ for polarised flux densities below 25 mJy. 

As a comparison with other methods for polarisation flux density estimations, the Bayesian method by \citet{HER21} provides reliable results down to 180 mJy when applied to realistic simulations with the QUIJOTE experiment characteristics. However, both the frequency analysed, namely 11 GHz, and the instrument sensitivity, 105 $\mu K s^{1/2}$, are different from our case (217 GHz with a $\sim$650 $\mu K s^{1/2}$ sensitivity). The Bayesian method performs similarly well to more traditional techniques such as the FF filter from \citet{ARG09}, which was used in \textit{Planck} \citep{PCCS2}.

\begin{figure}[ht]
\centering
\subfigure{\includegraphics[width=\linewidth]{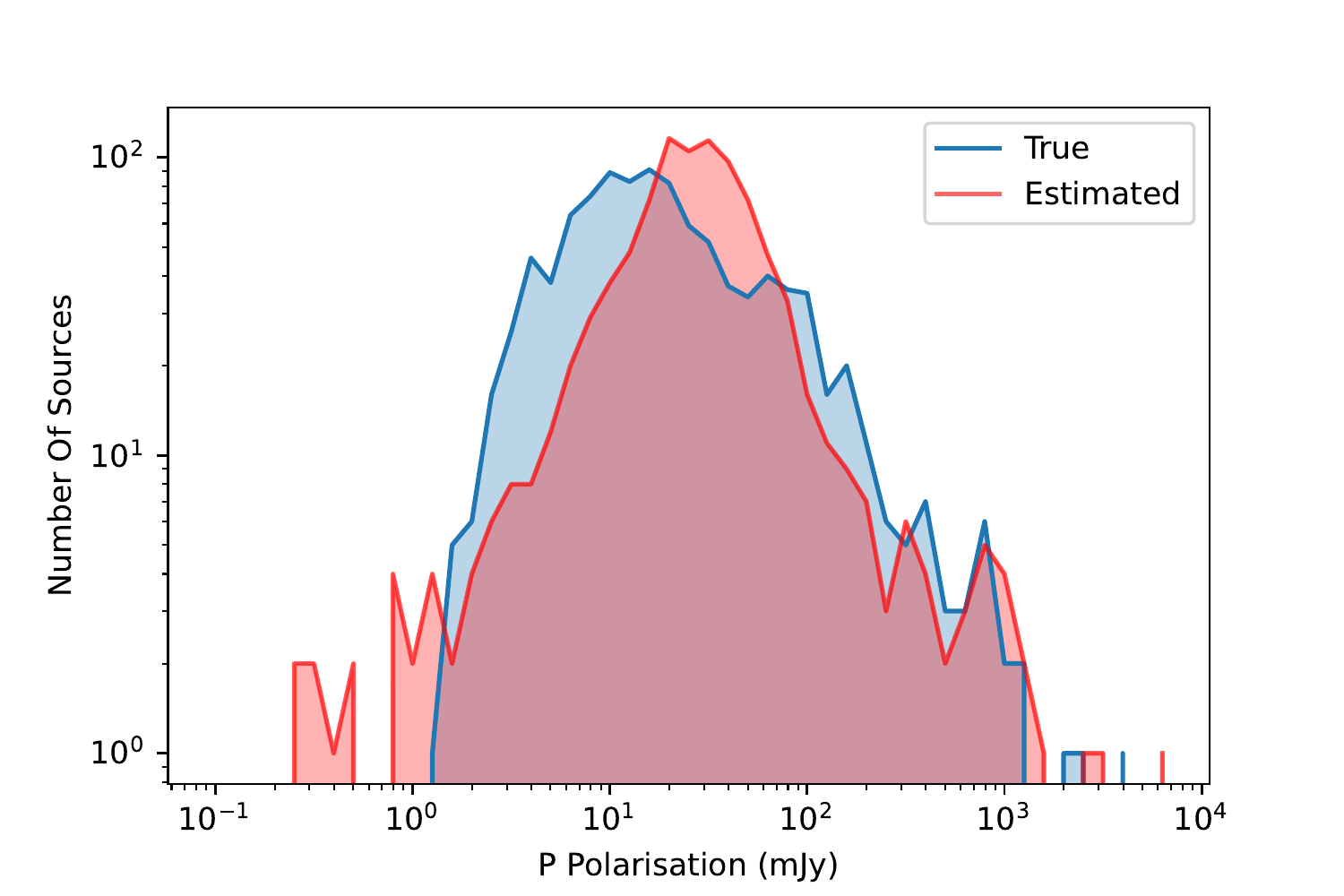}}
\caption{Source number counts in polarisation of both true (blue area) and estimated (red area) catalogues.}
\label{Fig:distributions_P}
\end{figure}

In Figure \ref{Fig:distributions_P}, we represent the true (blue area) and POSPEN-estimated (red area) source number counts in polarisation for the validation dataset. Both counts show a similar distribution, except in the 20-40 mJy flux density range, where the estimated counts are slightly higher with respect to the true ones. This is due to the Eddington bias, which also explains the fact that below 20 mJy, the estimated counts drop more rapidly than the true one.

\subsection{Polarisation angle estimation}

In this section, we follow a similar approach to that described in section~\ref{sec:results_P}, but with the aim of constraining the polarisation angle of the target PS. To do so, the CNN is re-trained, firstly for $Q$ and secondly for $U$ maps with 10000 simulated patches of PS and background, using the polarisation integrated values of $Q$ and $U$ central PS, respectively, to minimise the loss function. The CNN is then validated for each case with 1000 simulations not used for training.

\begin{figure*}[t]
\centering
\minipage{0.45\textwidth}
\includegraphics[width=\linewidth]{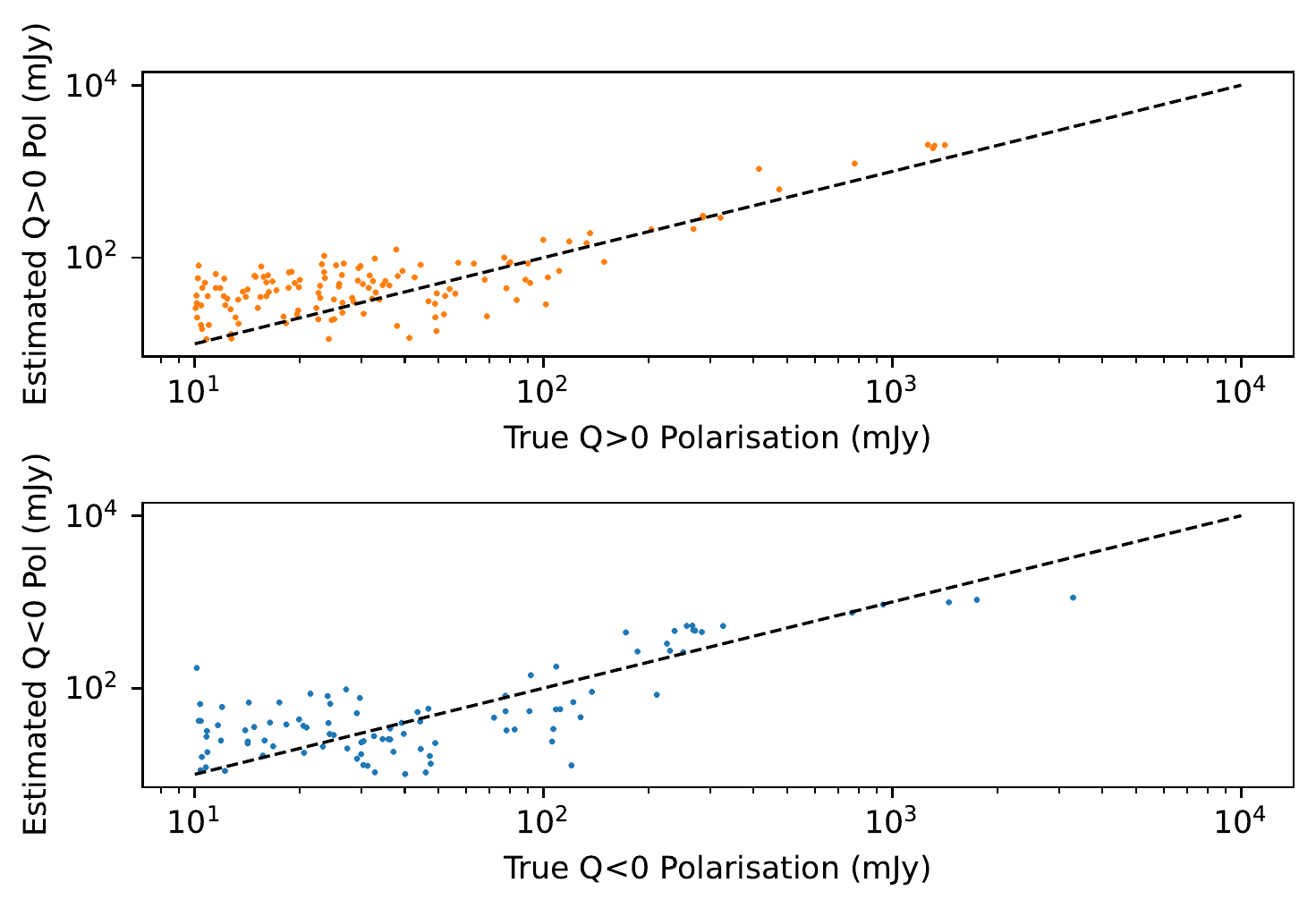}
\endminipage\hfill
\minipage{0.5\textwidth}%
  \includegraphics[width=\linewidth]{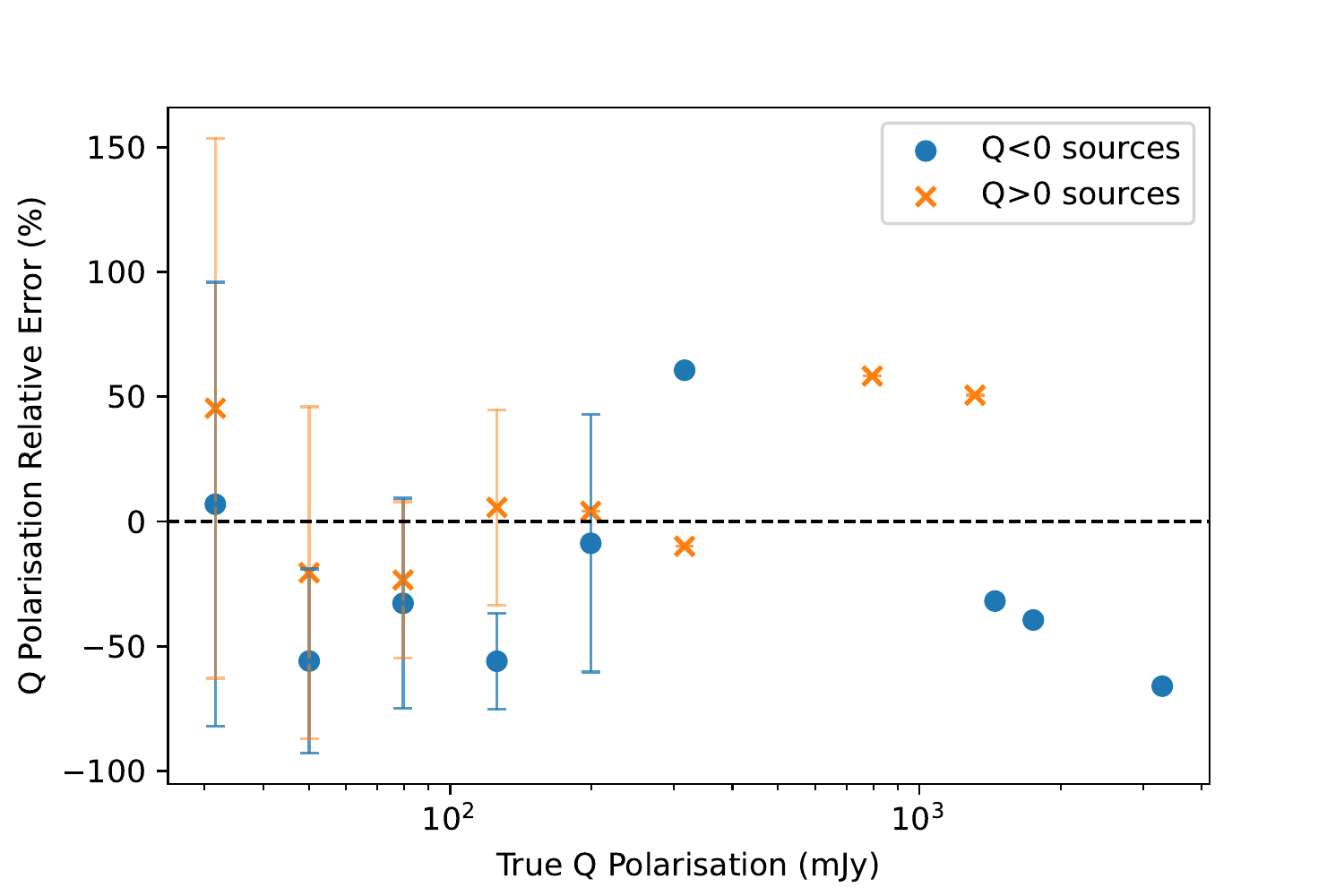}
\endminipage
\caption{Estimation of polarisation flux density of $Q$ sources. Left panel: Correlation between true and estimated $Q$ polarisation. The top part shows the $Q>0$ sources in orange while the bottom part shows the $Q<0$ ones in blue. Right panel: $Q$ relative error against the polarisation flux of the validation catalogue. The error bars represent the uncertainty computed as the standard deviation of each bin. In both cases, the dashed black line shows the ideal case.}
\label{Fig:true_vs_predicted_Q}
\end{figure*}

\begin{figure*}[t]
\centering
\minipage{0.45\textwidth}
\includegraphics[width=\linewidth]{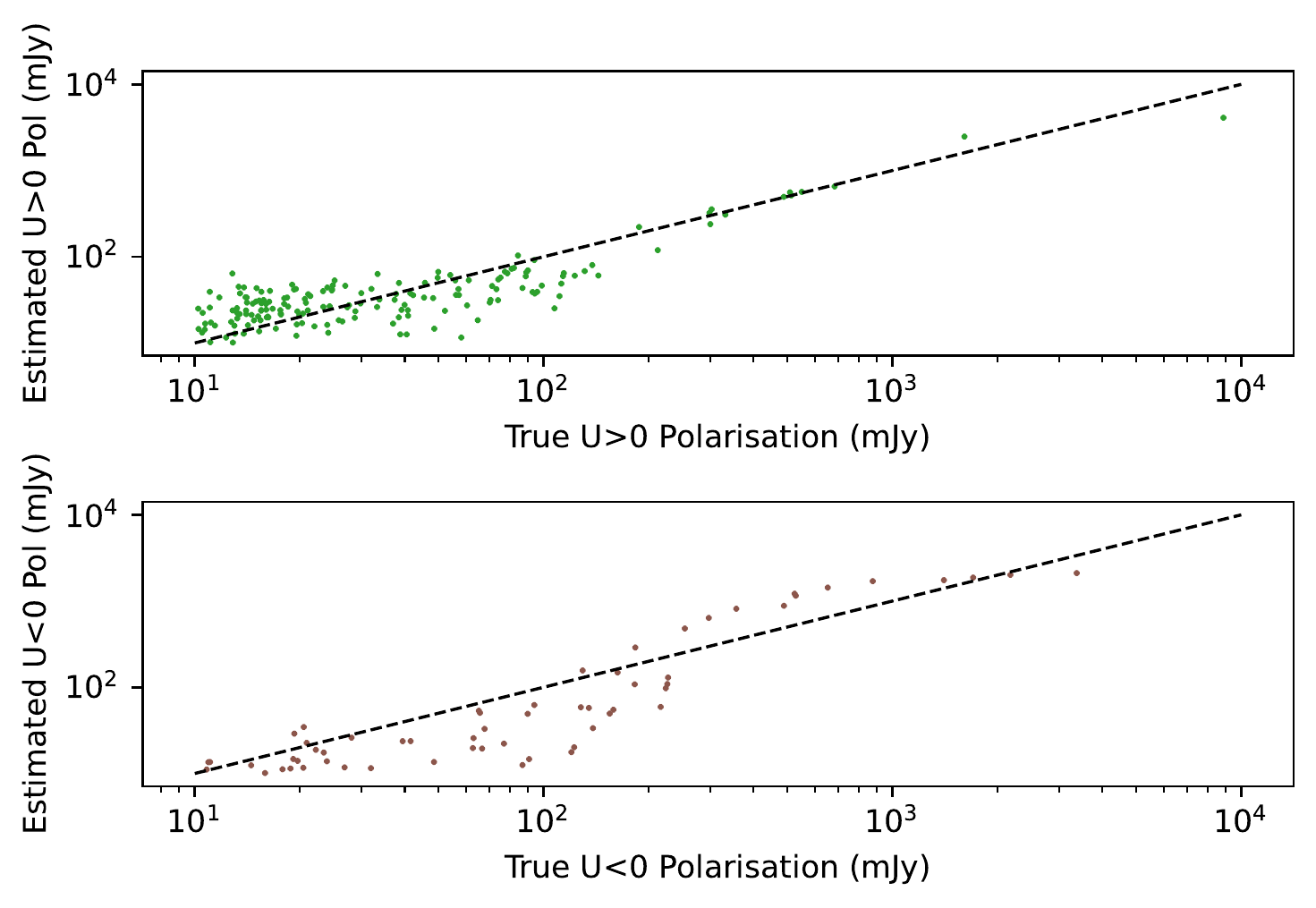}
\endminipage\hfill
\minipage{0.5\textwidth}%
  \includegraphics[width=\linewidth]{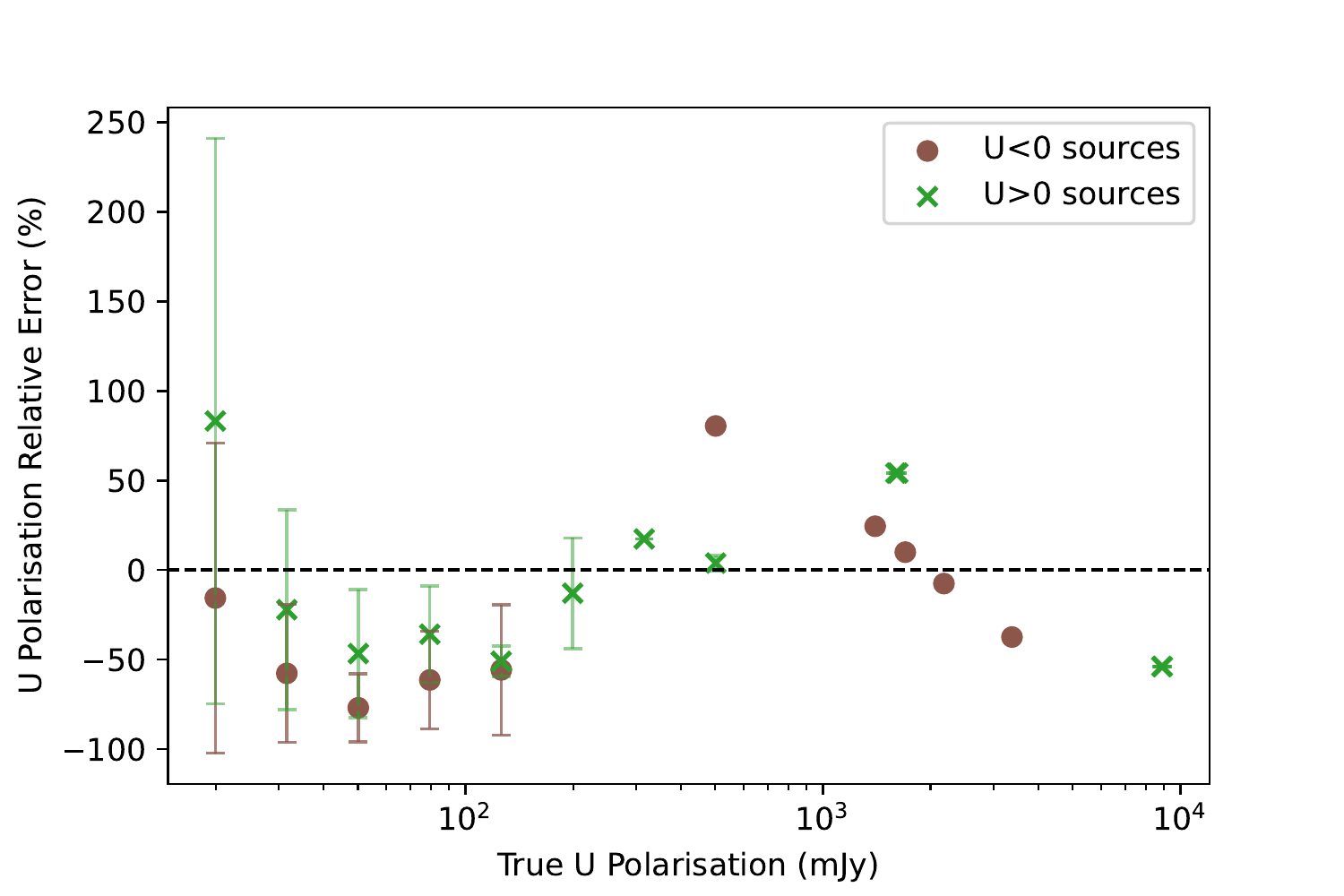}
\endminipage
\caption{Estimation of polarisation flux density of $U$ sources. Left panel: Correlation between true and estimated $U$ polarisation. The top part shows the $U>0$ sources in green while the bottom part shows the $U<0$ ones in brown. Right panel: $U$ relative error against the polarisation flux of the validation catalogue. The error bars represent the uncertainty computed as the standard deviation of each bin. In both cases, the dashed black line shows the ideal case.}
\label{Fig:true_vs_predicted_U}
\end{figure*}

The correlation between true and estimated $Q$ sources is shown in the left panel of Figure \ref{Fig:true_vs_predicted_Q}. The top panel represents sources with $Q>0$, that is, sources with a polarisation angle of between $0^\circ$ and $180^\circ$, while the bottom panel shows sources with $Q<0$, which are the ones with a polarisation angle between $-180^\circ$ and $0^\circ$. In the right panel, we represent the $Q$ relative error. The positive sources are represented as orange crosses and the negative as blue circles. In both cases, the uncertainty is taken as the standard deviation of each bin. For both figures, the black dashed lines represent the ideal case. The same statistical quantities for $U$ sources are plotted in Figure \ref{Fig:true_vs_predicted_U}.

In both cases, we can see a similar behaviour to the previous section: POSPEN recovers reliable sources above $\sim 80$ mJy. As expected, due to the fact that the network has to deal with positive and negative quantities, the errors in $Q$ and $U$ are higher than the ones in total polarisation $P$, reaching maximal relative errors of 60\% and 80\% for $Q$ and $U$ sources, respectively.

\begin{figure}[ht]
\centering
\subfigure{\includegraphics[width=\linewidth]{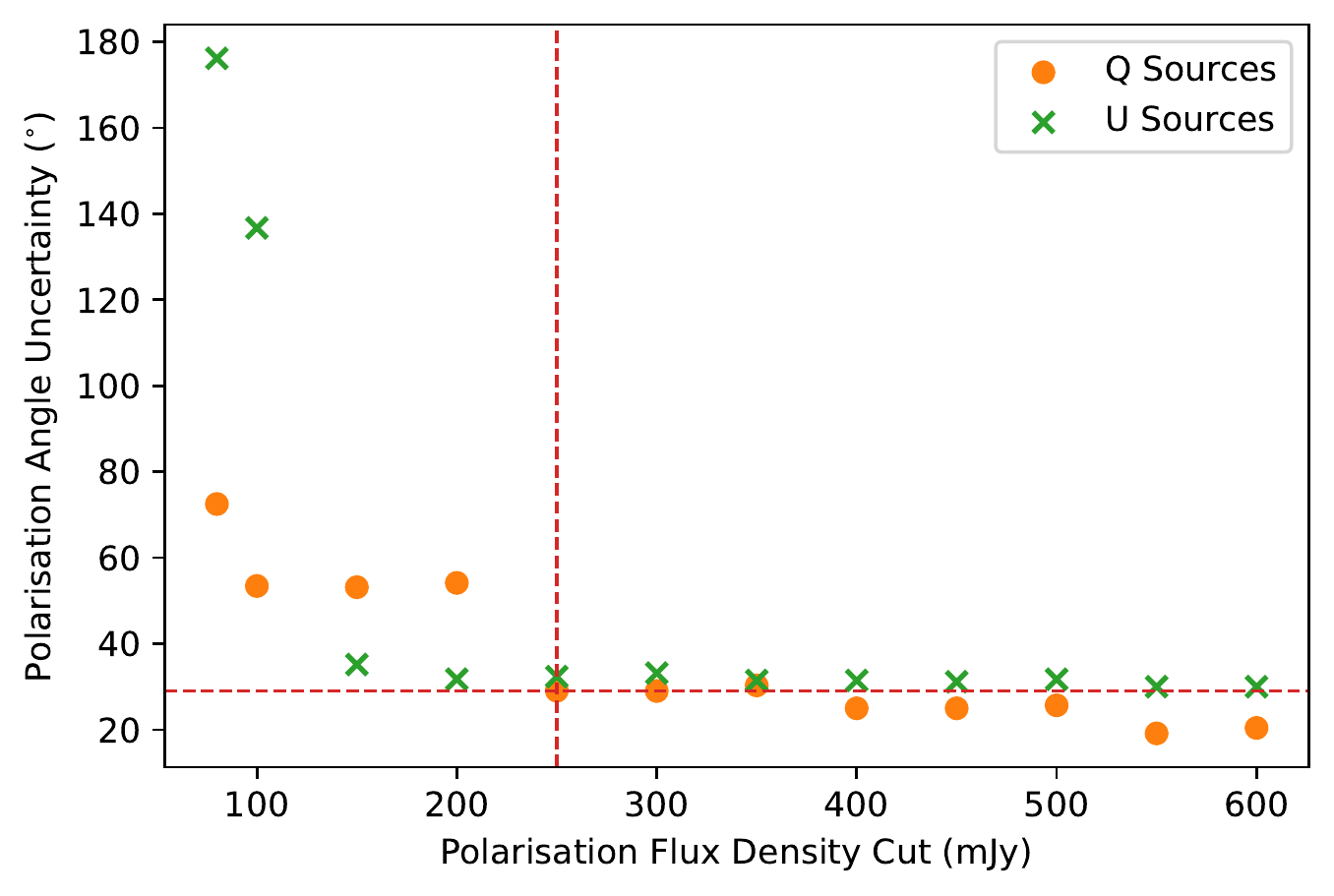}}
\caption{Relation between polarisation angle uncertainty for the estimations of POSPEN and the cut in polarisation flux density in both $Q$ and $U$ sources. The dashed red line shows the minimum uncertainty considering enough number of sources to have good statistics.}
\label{Fig:rms_vs_cut}
\end{figure}

Given these findings, in order to assess the performance of the CNN when estimating the polarisation angle, we consider only sources with reliable estimations above 80 mJy. We estimate the polarisation angle using eq. \eqref{eq:Q_U_expressions}, which is more suitable than eq. \eqref{eq:polarization_angle} because good-quality detections of both $Q$ and $U$ flux densities are not strictly required. With this approach, we can estimate the polarisation angle even when $Q$ is well estimated but not its corresponding $U$ value, or vice versa. 
However, in order to get the actual polarisation angle, we need to compare the $Q$ and $U$ values of every source: when $Q > 0$ and $U < 0,$ the actual angle is obtained by adding $\pi/2$ to the one from eq. \eqref{eq:Q_U_expressions}, while when $Q < 0$ and $U < 0$, we need to subtract $\pi/2$.

We also estimate the standard deviation of the relative error of the angle for each population in order to give the level of uncertainty of our method. This quantity depends on how well the polarisation flux of both $Q$ and $U$ sources is estimated. Therefore, to obtain the estimation for the polarisation angle with the minimum uncertainty, we examine the relation between the polarisation angle uncertainty and the polarised flux density cut for both $Q$ and $U$ sources, which we show in Figure \ref{Fig:rms_vs_cut}. We obtain that the minimal uncertainty that allows good statistics is $\pm$29$^{\circ}$ and $\pm$32$^{\circ}$ for a cut of $\sim$250 mJy for $Q$ and $U$ sources, respectively. Obviously, this cut could also be applied independently for each map in order to obtain estimations for sources with low $Q$ and high $U$ values, and vice versa.

\begin{figure*}[t]
\centering
\minipage{0.475\textwidth}
\includegraphics[width=8cm]{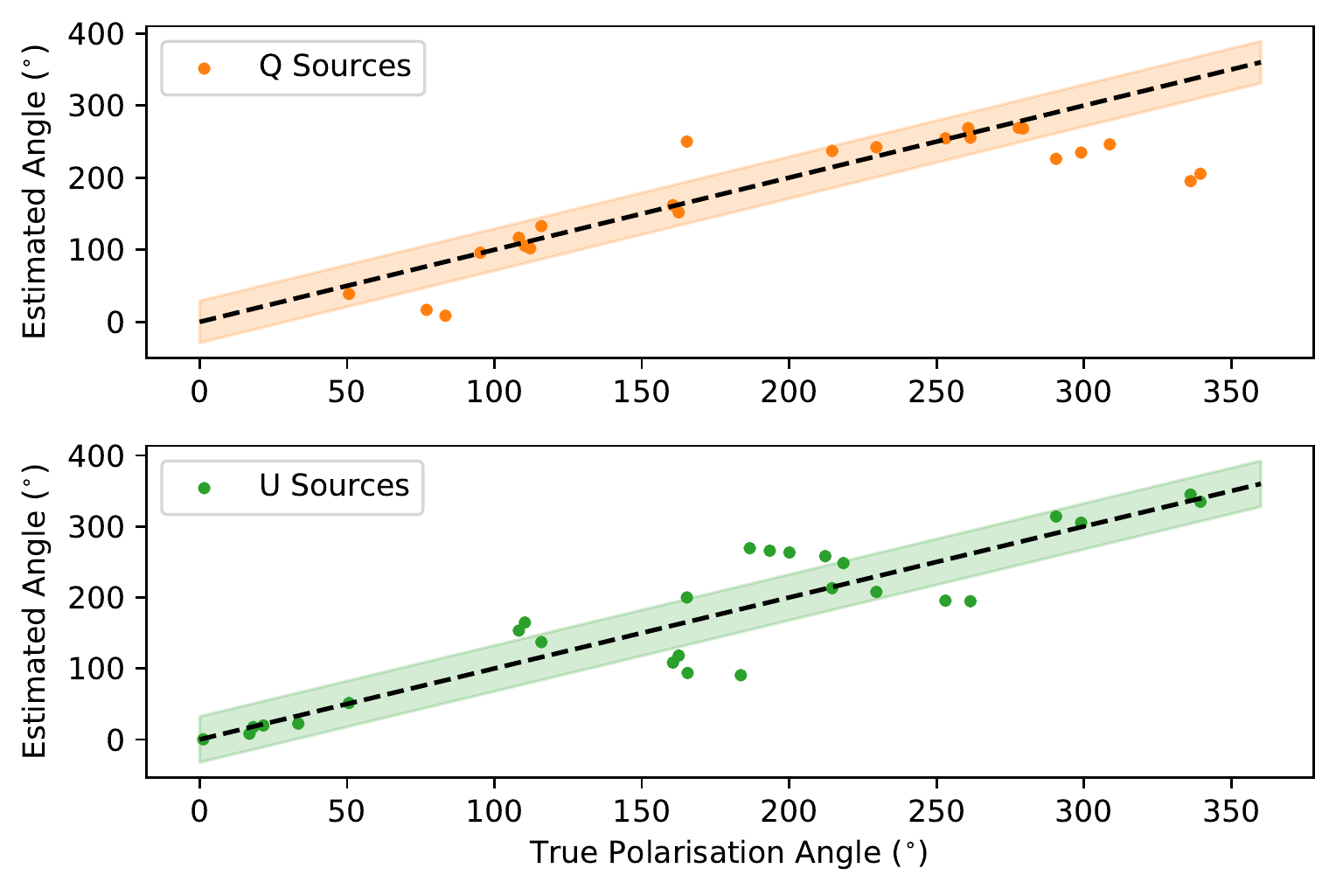}
\endminipage
\minipage{0.475\textwidth}%
  \includegraphics[width=8cm]{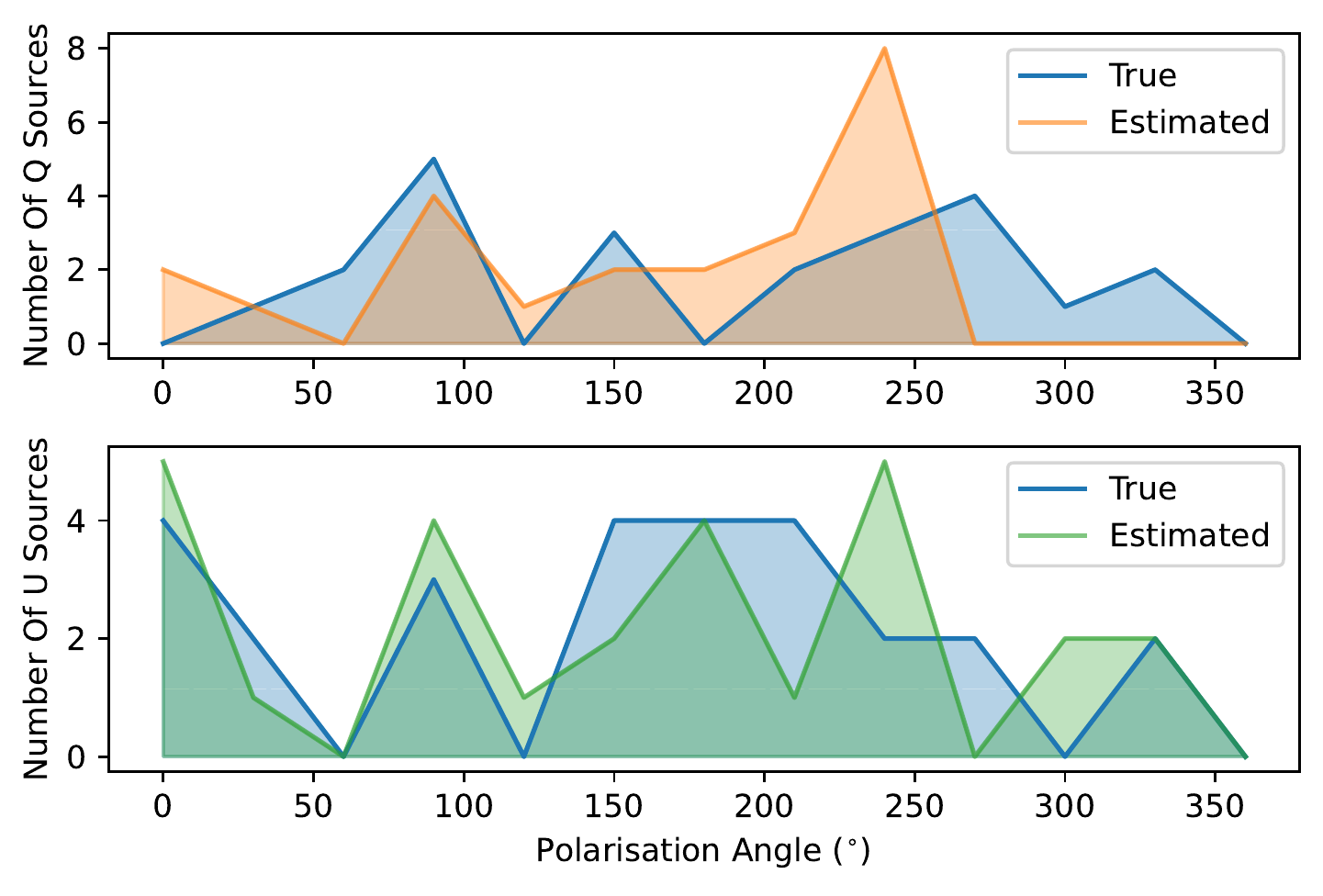}
\endminipage
\caption{Estimation of polarisation angles of $Q$ and $U$ sources. Left panel: Correlation between true and estimated angles. The coloured areas represent a confidence interval of $\pm29^{\circ}$ for $Q$ sources and $\pm32^{\circ}$ $U$ ones, which are the standard deviation of the relative error of each population, respectively. Dashed black lines are the 1:1 case. Right panel: Distribution of polarisation angles for both true and estimated $Q$ and $U$ catalogues, with the blue areas showing the true angles and orange and green areas showing the estimated angles for $Q$ and $U$ sources, respectively.}
\label{Fig:true_vs_predicted_angle}
\end{figure*}

Considering this cut, we show in the left panel of Figure \ref{Fig:true_vs_predicted_angle} the correlation between true and estimated polarisation angles for sources in $Q$ (top; orange) and $U$ (bottom; green), with the black dashed line showing the 1:1 case. The coloured areas represent the 1$\sigma$ uncertainty level computed as described above. As shown, most of the sources above the mentioned cut have their polarisation angles well recovered with an uncertainty of $\sim$30$^{\circ}$.

Furthermore, for assessing the quality of this correlation, we calculate the Pearson coefficient, defined as
\begin{equation}
    \rho_{\hspace{1pt} X, \hspace{1pt} Y} \hspace{1pt} = \hspace{1pt} \frac{cov \hspace{1pt} (X \hspace{1pt} Y)}{\sigma_{X} \hspace{1pt} \sigma_{Y}},
\label{eq:mse}
\end{equation}
where $cov \hspace{1pt} (X \hspace{2pt} Y)$ is the covariance between true and estimated angles and $\sigma_{X}$ and $\sigma_{Y}$ are their standard deviations. We obtain a Pearson coefficient of 0.853 for $Q$ sources and 0.91 for $U$ ones. We also implement a test of null hypothesis, obtaining a probability value of 5.87$\times 10^{-6}$\% for $Q$ sources and 1.03$\times 10^{-10}$\% for $U$ ones.

In the right panels of Figure \ref{Fig:true_vs_predicted_angle}, we plot the distribution of the input angles (blue area in both top and bottom panels) and $Q$ (top; orange area) and $U$ (bottom; green area) estimated ones. This shows on the one hand that the input sources have relatively random angles, as expected, and on the other hand that POSPEN estimates the distribution of the sources in most of the intervals  relatively well. 

\subsection{Comparison with the PCCS2}

In this section, we apply our model to the 217 GHz \textit{Planck} data. In particular, we extract patches of  32$\times$32 pixels in area from the total 217 GHz \textit{Planck} map. The patches are centred on the positions of the PCCS2 sources with polarisation detection. Thus, the resulting dataset is made up of 11 patches. We apply POSPEN to this dataset and obtain estimations of the corresponding total polarisation flux densities and angles. We compare our results against the PCCS2 estimated values.

\begin{figure*}[t]
\centering
\minipage{0.5\textwidth}
\includegraphics[width=\linewidth]{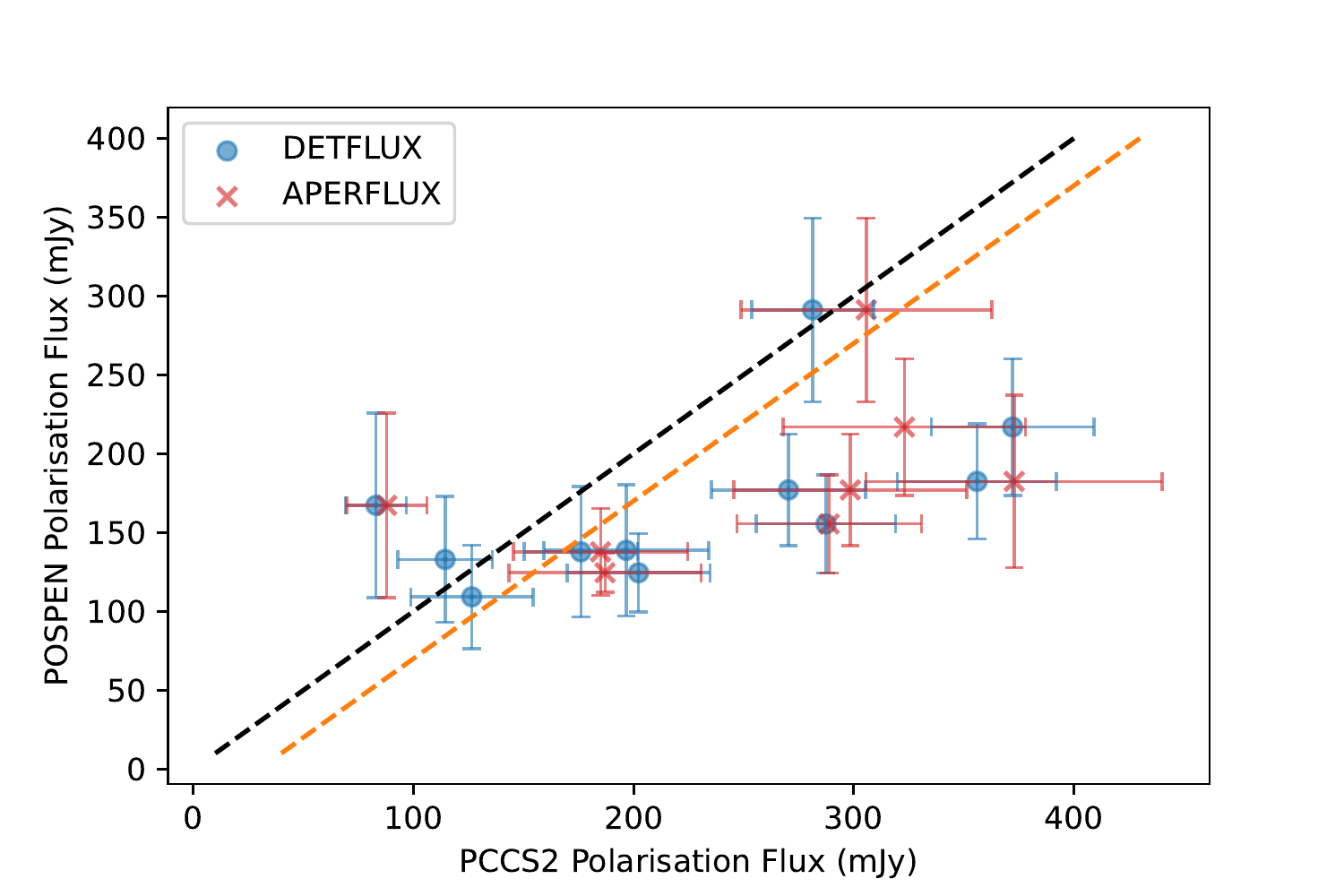}
\endminipage\hfill
\minipage{0.5\textwidth}%
  \includegraphics[width=\linewidth]{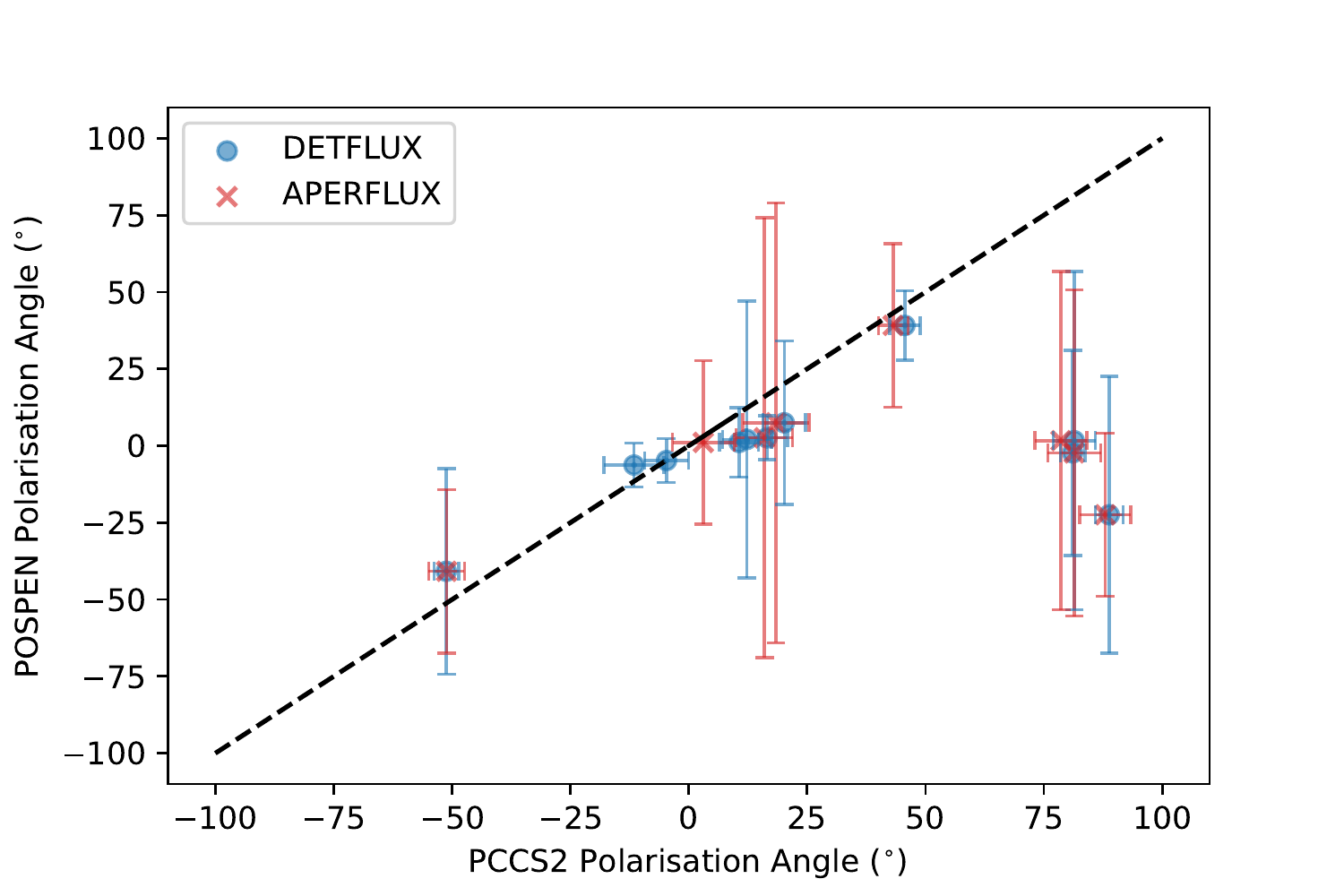}
\endminipage
\caption{Comparison of polarisation flux densities and angle estimations given by both POSPEN and the PCCS2. Left panel: Comparison of PCCS2 with POSPEN polarisation flux density estimations. Right panel: Comparison of PCCS2 with POSPEN polarisation angle estimations. In both cases, the horizontal error bars show the uncertainty given in the PCCS2, while the vertical ones show the uncertainty of POSPEN. The dashed orange line shows the systematic underestimation of POSPEN given by the simulations and the dashed black line is the 1:1 case.}
\label{Fig:comparison_PCCS2}
\end{figure*}

Figure \ref{Fig:comparison_PCCS2} shows the comparison between the estimations given in the PCCS2 by using the detection flux (DETFLUX, blue points) and aperture flux (APERFLUX, red crosses) methods\footnote{These are extensively described in \citet{PCCS}} and the estimations by POSPEN: the left panel compares the total polarisation flux densities and the right panel the polarisation angles. The corresponding horizontal error bars are the uncertainty given in the PCCS2, the vertical ones are the uncertainty of POSPEN, and the dashed black lines are the 1:1 case. The results obtained with simulations (described in the previous subsections) are used for assigning the POSPEN uncertainties to both polarisation flux densities and angles. More specifically, we use the relative error shown in the right panel of Figure \ref{Fig:correlation_P} for each flux density interval to assign the uncertainty to each source according to its polarisation level. The same procedure is applied to the polarisation angle, using the uncertainties shown in the right panels of Figures \ref{Fig:true_vs_predicted_Q} and \ref{Fig:true_vs_predicted_U}.

Regarding the total polarisation, both for DETFLUX and APERTURE values, POSPEN tends to underestimate the flux densities of six of the detected PCCS2 sources, while overestimating the faintest one. Furthermore, all these sources were considered to be robustly detected in \textit{Planck} and are relatively bright in polarisation, with a minimum polarisation of about 80 mJy. These results could simply indicate a potential issue with POSPEN in this polarisation flux density range. Indeed, a statistical underestimation can be seen in the 200-300 mJy range in the right panel of Figure \ref{Fig:correlation_P}. The simulated sources with polarisation flux density brighter than 100 mJy constitute only $\sim$20\% of the whole sample. 
In fact, although we obtain interesting results, the fluctuations in the relative errors above 80 mJy that we see in Figure \ref{Fig:correlation_P} indicate that the training dataset has to be improved. We must also take into account the fact that, in real data, highly polarised sources are usually detected in polarisation maps, with it being very difficult to single out those sources with flux densities lower than 100 mJy. Our training set contains 10000 sources and only a few have flux densities above 100 mJy. Therefore, it might be worth injecting brighter sources into the training set when validating on real data.

Therefore, the performance of POSPEN could probably be improved when validating the hole PCCS2 catalogue by increasing the number of simulated sources brighter than 100 mJy, which are the most relevant ones when applied to the real maps. 
However, this is out of the scope of this work and will be tested in the future.

 The polarisation angle is only underestimated for three sources by POSPEN in both DETFLUX and APERFLUX measurements, and the other eight estimations are consistent with \textit{Planck} values. Therefore, although most of the results in total polarisation flux density underestimate \textit{Planck} values, the polarisation angle obtained using $Q$ and $U$ data are mostly in agreement. 

Finally, we also estimate the fractional polarisation, $\Pi_{estimated} \hspace{1pt} = \hspace{1pt} \frac{P_{estimated}}{S_{true}} \hspace{1pt}$, of the sources detected in the PCCS2. We obtain a mean fractional polarisation of 4.78$\%$, a value somewhat lower than the 5.3$\%$ in \citet{PCCS2}, and a median of 2.82$\%$, which is similar to the findings of other works in the literature that provide estimations using single-channel \textit{Planck} data \citep[as in][]{TRO18}.


\section{Conclusions}
\label{sec:conclusions}

In this work, we aim to develop a new non-blind methodology for point-source polarisation flux density and angle estimation based on a CNN called the POint Source Polarisation Estimation Network (POSPEN). The CNN is trained with 10000 realistic simulations of $32\times32$ pixel patches (pixel size of 90 arcsec) composed of a central injected PS embedded in a realistic background with the 217 GHz \textit{Planck} channel characteristics. The PSs are simulated following the C2Ex model by \citet{TUC11}, and their polarisation is simulated assuming a log-normal distribution with $\mu$ and $\sigma$ parameters from \citet{Bon17a}. The background is composed of a simulated map of Galactic thermal dust, the cosmic microwave background map from the \texttt{SEVEM} method, and injected instrumental white noise using the \textit{Planck} value.

POSPEN is then validated with simulations not used for training, splitting our analysis into three parts. Firstly, we study the performance of the CNN on total polarisation data.
We compute the correlation between true and estimated polarisation flux densities along with the relative error with respect to the input polarisation flux densities. We obtain that our method reliably recovers the polarisation flux density of sources above 80 mJy, where the Eddington bias starts to affect the results. Under these circumstances, we are able to recover the input polarisation flux density within a relative error of 30\% in most of the flux-density intervals. 

Subsequently, we re-train the neural network with 10000 simulations in both $Q$ and $U$ maps in order to obtain a constraint on the polarisation angle. We study the correlation between true and estimated $Q$ and $U$ values and their errors. We obtain similar results to in the previous case, reaching the Eddington bias at approximately the same flux-density level. For constraining the polarisation angle, we then use sources above this flux-density limit, obtaining minimal 1$\sigma$ uncertainties of $\pm$29$^{\circ}$ and $\pm$32$^{\circ}$ for $Q$ and $U$ sources, respectively, above 250 mJy. For these, we analyse the correlation between these true and estimated angles and their distributions. We obtain a Pearson coefficient of 0.853 for $Q$ sources and 0.91 for $U$ sources with null-hypothesis probability values of 5.87$\times$10$^{-6}$\% and 1.03$\times$10$^{-10}$\%, respectively. 

Finally, we study the performance of our model with real data from the 217 GHz \textit{Planck} map and compare our estimations with the information in the PCCS2. We find that POSPEN recovers a similar polarisation flux density for 4 of the 11 sources in the PCCS2, but it tends to systematically underestimate the polarisation flux density of 6 sources and overestimate the value of the remaining one. For the polarisation angle, it recovers this quantity quite well for 8 sources, underestimating 3 of them. Finally, we compute the mean and median fractional polarisation values, obtaining $4.78\%$ and $2.82\%$, respectively.

Based on these results, the methodology presented here appears to be promising for estimating polarisation flux density and angle in a non-blind way, especially considering the level of contamination in the simulations, the data we use (mimicking the \textit{Planck} experiment), and the performance of other non-blind methods in similar conditions. In particular, the results that can be obtained by applying this method to real data can be useful not only for constraining the impact of polarised sources on the detection of primordial B-modes ---if the tensor-to-scalar ratio is lower than r = 0.001--- but also for improving our knowledge of active galactic nuclei.

However, when considering the application of this method in conjunction with other instruments, in order to obtain a good performance, the network should be retrained with a training set consisting of realistic simulations that reproduce the characteristics and conditions of such instruments, such as the higher resolution ACTPol or SPTPol experiments. Moreover, this method can be further improved in order to detect sources in polarisation data in a blind way by subsequently applying the fully convolutional neural network by \citet{CAS22} to perform the detection in total intensity and then the methodology presented in this work to estimate the polarisation flux density and angle of the detected sources.

\begin{acknowledgements}
We warmly thank the anonymous referee for the very useful comments on the original manuscript. JMC, LB, JGN, MMC and DC acknowledge financial support from the PID2021-125630NB-I00 project funded by MCIN/AEI/10.13039/501100011033 / FEDER, UE. JMC also acknowledges financial support from the SV-PA-21-AYUD/2021/51301 project. MMC also acknowledges to be granted by PAPI-21-PF-04 (Universidad de Oviedo). CGC, JDS, MLS and FJDC acknowledge financial support from both PID2021-127331NB-I00 and UE-18-SOLARNET-824135 projects. \\
The authors thank Prof. José Alberto Rubiño-Martín, Prof. Ricardo Tanasú Génova-Santos for valuable comments. They also thank Sergio Ena for helping with polarisation fraction results at the first stages of the original manuscript. \\
This research has made use of the python packages \texttt{Matplotlib} \citep{matplotlib}, \texttt{Pandas} \citep{pandas}, \texttt{Keras} \citep{KER}, and \texttt{Numpy} \citep{numpy}, also the \texttt{HEALPix} \citep{GOR05} and \texttt{Healpy} \citep{zon19} packages.
\end{acknowledgements}

%
%
\bibliographystyle{aa}
\bibliography{SDNN}

\begin{thebibliography}{47}
\expandafter\ifx\csname natexlab\endcsname\relax\def\natexlab#1{#1}\fi

\bibitem[{{Arg{\"u}eso} {et~al.}(2009){Arg{\"u}eso}, {Sanz}, {Herranz},
  {L{\'o}pez-Caniego}, \& {Gonz{\'a}lez-Nuevo}}]{ARG09}
{Arg{\"u}eso}, F., {Sanz}, J.~L., {Herranz}, D., {L{\'o}pez-Caniego}, M., \&
  {Gonz{\'a}lez-Nuevo}, J. 2009, \mnras, 395, 649

\bibitem[{{Bonavera} {et~al.}(2017{\natexlab{a}}){Bonavera},
  {Gonz{\'a}lez-Nuevo}, {Arg{\"u}eso}, \& {Toffolatti}}]{Bon17a}
{Bonavera}, L., {Gonz{\'a}lez-Nuevo}, J., {Arg{\"u}eso}, F., \& {Toffolatti},
  L. 2017{\natexlab{a}}, \mnras, 469, 2401

\bibitem[{{Bonavera} {et~al.}(2017{\natexlab{b}}){Bonavera},
  {Gonz{\'a}lez-Nuevo}, {De Marco}, {Arg{\"u}eso}, \& {Toffolatti}}]{Bon17b}
{Bonavera}, L., {Gonz{\'a}lez-Nuevo}, J., {De Marco}, B., {Arg{\"u}eso}, F., \&
  {Toffolatti}, L. 2017{\natexlab{b}}, \mnras, 472, 628

\bibitem[{{Bonavera} {et~al.}(2021){Bonavera}, {Suarez Gomez},
  {Gonz{\'a}lez-Nuevo}, {Cueli}, {Santos}, {Sanchez}, {Mu{\~n}iz}, \& {de
  Cos}}]{BON21}
{Bonavera}, L., {Suarez Gomez}, S.~L., {Gonz{\'a}lez-Nuevo}, J., {et~al.} 2021,
  \aap, 648, A50

\bibitem[{{Casas} {et~al.}(2022{\natexlab{a}}){Casas}, {Bonavera},
  {Gonz{\'a}lez-Nuevo}, {Baccigalupi}, {Cueli}, {Crespo}, {Goitia}, {Santos},
  {S{\'a}nchez}, \& {de Cos}}]{CAS22b}
{Casas}, J.~M., {Bonavera}, L., {Gonz{\'a}lez-Nuevo}, J., {et~al.}
  2022{\natexlab{a}}, \aap, 666, A89

\bibitem[{{Casas} {et~al.}(2022{\natexlab{b}}){Casas}, {Gonz{\'a}lez-Nuevo},
  {Bonavera}, {Herranz}, {Suarez Gomez}, {Cueli}, {Crespo}, {Santos},
  {S{\'a}nchez}, {S{\'a}nchez-Lasheras}, \& {de Cos}}]{CAS22}
{Casas}, J.~M., {Gonz{\'a}lez-Nuevo}, J., {Bonavera}, L., {et~al.}
  2022{\natexlab{b}}, \aap, 658, A110

\bibitem[{Chollet(2015)}]{KER}
Chollet, F. 2015, Keras, \url{https://github.com/fchollet/keras}

\bibitem[{{Datta} {et~al.}(2019){Datta}, {Aiola}, {Choi}, {Devlin}, {Dunkley},
  {D{\"u}nner}, {Gallardo}, {Gralla}, {Halpern}, {Hasselfield}, {Hilton},
  {Hincks}, {Ho}, {Hubmayr}, {Huffenberger}, {Hughes}, {Kosowsky},
  {L{\'o}pez-Caraballo}, {Louis}, {Lungu}, {Marriage}, {Maurin}, {McMahon},
  {Moodley}, {Naess}, {Nati}, {Niemack}, {Page}, {Partridge}, {Prince},
  {Staggs}, {Switzer}, {Wollack}, \& {Farren}}]{ACT19}
{Datta}, R., {Aiola}, S., {Choi}, S.~K., {et~al.} 2019, \mnras, 486, 5239

\bibitem[{{de Zotti} {et~al.}(2010){de Zotti}, {Massardi}, {Negrello}, \&
  {Wall}}]{deZOT09}
{de Zotti}, G., {Massardi}, M., {Negrello}, M., \& {Wall}, J. 2010, \aapr, 18,
  1

\bibitem[{{Delabrouille} {et~al.}(2013){Delabrouille}, {Betoule}, {Melin},
  {Miville-Desch{\^e}nes}, {Gonzalez-Nuevo}, {Le Jeune}, {Castex}, {de Zotti},
  {Basak}, {Ashdown}, {Aumont}, {Baccigalupi}, {Band ay}, {Bernard}, {Bouchet},
  {Clements}, {da Silva}, {Dickinson}, {Dodu}, {Dolag}, {Elsner}, {Fauvet},
  {Fa{\"y}}, {Giardino}, {Leach}, {Lesgourgues}, {Liguori},
  {Mac{\'\i}as-P{\'e}rez}, {Massardi}, {Matarrese}, {Mazzotta}, {Montier},
  {Mottet}, {Paladini}, {Partridge}, {Piffaretti}, {Prezeau}, {Prunet},
  {Ricciardi}, {Roman}, {Schaefer}, \& {Toffolatti}}]{Del13}
{Delabrouille}, J., {Betoule}, M., {Melin}, J.~B., {et~al.} 2013, \aap, 553,
  A96

\bibitem[{{Diego-Palazuelos} {et~al.}(2021){Diego-Palazuelos}, {Vielva}, \&
  {Herranz}}]{DP21}
{Diego-Palazuelos}, P., {Vielva}, P., \& {Herranz}, D. 2021, \jcap, 2021, 048

\bibitem[{{Galluzzi} {et~al.}(2018){Galluzzi}, {Massardi}, {Bonaldi},
  {Casasola}, {Gregorini}, {Trombetti}, {Burigana}, {Bonato}, {De Zotti},
  {Ricci}, {Stevens}, {Ekers}, {Bonavera}, {di Serego Alighieri}, {Liuzzo},
  {L{\'o}pez-Caniego}, {Paladino}, {Toffolatti}, {Tucci}, \&
  {Callingham}}]{GAL18}
{Galluzzi}, V., {Massardi}, M., {Bonaldi}, A., {et~al.} 2018, \mnras, 475, 1306

\bibitem[{{Gonz{\'a}lez-Nuevo} {et~al.}(2006){Gonz{\'a}lez-Nuevo},
  {Arg{\"u}eso}, {L{\'o}pez-Caniego}, {Toffolatti}, {Sanz}, {Vielva}, \&
  {Herranz}}]{GN06}
{Gonz{\'a}lez-Nuevo}, J., {Arg{\"u}eso}, F., {L{\'o}pez-Caniego}, M., {et~al.}
  2006, \mnras, 369, 1603

\bibitem[{{Gonz{\'a}lez-Nuevo} {et~al.}(2005){Gonz{\'a}lez-Nuevo},
  {Toffolatti}, \& {Arg{\"u}eso}}]{GN05}
{Gonz{\'a}lez-Nuevo}, J., {Toffolatti}, L., \& {Arg{\"u}eso}, F. 2005, \apj,
  621, 1

\bibitem[{Goodfellow {et~al.}(2016)Goodfellow, Bengio, \& Courville}]{GOO16}
Goodfellow, I.~J., Bengio, Y., \& Courville, A. 2016, Deep Learning (Cambridge,
  MA, USA: MIT Press), \url{http://www.deeplearningbook.org}

\bibitem[{{G{\'o}rski} {et~al.}(2005){G{\'o}rski}, {Hivon}, {Banday}, {Wand
  elt}, {Hansen}, {Reinecke}, \& {Bartelmann}}]{GOR05}
{G{\'o}rski}, K.~M., {Hivon}, E., {Banday}, A.~J., {et~al.} 2005, \apj, 622,
  759

\bibitem[{{Gupta} {et~al.}(2019){Gupta}, {Reichardt}, {Ade}, {Anderson},
  {Archipley}, {Austermann}, {Avva}, {Beall}, {Bender}, {Benson}, {Bianchini},
  {Bleem}, {Carlstrom}, {Chang}, {Chiang}, {Citron}, {Moran}, {Crawford},
  {Crites}, {de Haan}, {Dobbs}, {Everett}, {Feng}, {Gallicchio}, {George},
  {Gilbert}, {Halverson}, {Harrington}, {Henning}, {Hilton}, {Holder},
  {Holzapfel}, {Hou}, {Hrubes}, {Huang}, {Hubmayr}, {Irwin}, {Knox}, {Lee},
  {Li}, {Lowitz}, {Luong-Van}, {Marrone}, {McMahon}, {Meyer}, {Mocanu}, {Mohr},
  {Montgomery}, {Nadolski}, {Natoli}, {Nibarger}, {Noble}, {Novosad}, {Padin},
  {Patil}, {Pryke}, {Ruhl}, {Saliwanchik}, {Sayre}, {Schaffer}, {Shirokoff},
  {Sievers}, {Smecher}, {Staniszewski}, {Stark}, {Story}, {Switzer}, {Tucker},
  {Vanderlinde}, {Veach}, {Vieira}, {Wang}, {Whitehorn}, {Williamson}, {Wu},
  {Yefremenko}, \& {Zhang}}]{SPT19}
{Gupta}, N., {Reichardt}, C.~L., {Ade}, P.~A.~R., {et~al.} 2019, \mnras, 490,
  5712

\bibitem[{{Hamaker, J. P.} \& {Bregman, J. D.}(1996)}]{HAM96}
{Hamaker, J. P.} \& {Bregman, J. D.} 1996, Astron. Astrophys. Suppl. Ser., 117,
  161

\bibitem[{{Hanany} {et~al.}(2019){Hanany}, {Alvarez}, {Artis}, {Ashton},
  {Aumont}, {Aurlien}, {Banerji}, {Barreiro}, {Bartlett}, {Basak}, {Battaglia},
  {Bock}, {Boddy}, {Bonato}, {Borrill}, {Bouchet}, {Boulanger}, {Burkhart},
  {Chluba}, {Chuss}, {Clark}, {Cooperrider}, {Crill}, {De Zotti},
  {Delabrouille}, {Di Valentino}, {Didier}, {Dor{\'e}}, {Eriksen}, {Errard},
  {Essinger-Hileman}, {Feeney}, {Filippini}, {Fissel}, {Flauger}, {Fuskeland},
  {Gluscevic}, {Gorski}, {Green}, {Hensley}, {Herranz}, {Hill}, {Hivon},
  {Hlo{\v{z}}ek}, {Hubmayr}, {Johnson}, {Jones}, {Jones}, {Knox}, {Kogut},
  {L{\'o}pez-Caniego}, {Lawrence}, {Lazarian}, {Li}, {Madhavacheril}, {Melin},
  {Meyers}, {Murray}, {Negrello}, {Novak}, {O'Brient}, {Paine}, {Pearson},
  {Pogosian}, {Pryke}, {Puglisi}, {Remazeilles}, {Rocha}, {Schmittfull},
  {Scott}, {Shirron}, {Stephens}, {Sutin}, {Tomasi}, {Trangsrud}, {van
  Engelen}, {Vansyngel}, {Wehus}, {Wen}, {Xu}, {Young}, \& {Zonca}}]{Han19}
{Hanany}, S., {Alvarez}, M., {Artis}, E., {et~al.} 2019, arXiv e-prints,
  arXiv:1902.10541

\bibitem[{Herranz {et~al.}(2012)Herranz, Argueso, \& Carvalho}]{HER12}
Herranz, D., Argueso, F., \& Carvalho, P. 2012, Advances in Astronomy, 2012,
  410965

\bibitem[{{Herranz} {et~al.}(2021){Herranz}, {Arg\"ueso, F.}, {Toffolatti, L.},
  {Manj\'on-Garc\'{\i}a, A.}, \& {L\'opez-Caniego, M.}}]{HER21}
{Herranz}, D., {Arg\"ueso, F.}, {Toffolatti, L.}, {Manj\'on-Garc\'{\i}a, A.},
  \& {L\'opez-Caniego, M.} 2021, A\&A, 651, A24

\bibitem[{{Herranz} {et~al.}(2009){Herranz}, {L{\'o}pez-Caniego}, {Sanz}, \&
  {Gonz{\'a}lez-Nuevo}}]{Her09}
{Herranz}, D., {L{\'o}pez-Caniego}, M., {Sanz}, J.~L., \& {Gonz{\'a}lez-Nuevo},
  J. 2009, \mnras, 394, 510

\bibitem[{Hunter(2007)}]{matplotlib}
Hunter, J.~D. 2007, Computing In Science \& Engineering, 9, 90

\bibitem[{{Krachmalnicoff} \& {Puglisi}(2021)}]{KRA21}
{Krachmalnicoff}, N. \& {Puglisi}, G. 2021, \apj, 911, 42

\bibitem[{{Krachmalnicoff} \& {Tomasi}(2019)}]{KRA19}
{Krachmalnicoff}, N. \& {Tomasi}, M. 2019, \aap, 628, A129

\bibitem[{LeCun {et~al.}(1989)LeCun, Boser, Denker, Henderson, Howard, Hubbard,
  \& Jackel}]{LeC89}
LeCun, Y., Boser, B., Denker, J.~S., {et~al.} 1989, Neural Computation, 1, 541

\bibitem[{Linde(1982)}]{LIN82}
Linde, A. 1982, Physics Letters B, 108, 389

\bibitem[{{L{\'o}pez-Caniego} {et~al.}(2009){L{\'o}pez-Caniego}, {Massardi},
  {Gonz{\'a}lez-Nuevo}, {Lanz}, {Herranz}, {De Zotti}, {Sanz}, \&
  {Arg{\"u}eso}}]{LOP09}
{L{\'o}pez-Caniego}, M., {Massardi}, M., {Gonz{\'a}lez-Nuevo}, J., {et~al.}
  2009, \apj, 705, 868

\bibitem[{Martínez-González {et~al.}(2003)Martínez-González, Diego, Vielva,
  \& Silk}]{MartinezGonzalez2003}
Martínez-González, E., Diego, J.~M., Vielva, P., \& Silk, J. 2003, Monthly
  Notices of the Royal Astronomical Society, 345, 1101

\bibitem[{Nair \& Hinton(2010)}]{NAI10}
Nair, V. \& Hinton, G.~E. 2010, in Proceedings of the 27th international
  conference on machine learning (ICML-10), 807--814

\bibitem[{Oliphant(2006)}]{numpy}
Oliphant, T. 2006, {NumPy}: A guide to {NumPy}, USA: Trelgol Publishing,
  [Online; accessed <today>]

\bibitem[{{Pearson} \& {Readhead}(1984)}]{PEA84}
{Pearson}, T.~J. \& {Readhead}, A.~C.~S. 1984, \araa, 22, 97

\bibitem[{{Planck Collaboration I}(2020)}]{PLA_18_I}
{Planck Collaboration I}. 2020, \aap, 641, A1

\bibitem[{{Planck Collaboration XXVI}(2016)}]{PCCS2}
{Planck Collaboration XXVI}. 2016, \aap, 594, A26

\bibitem[{{Planck Collaboration XXVIII}(2014)}]{PCCS}
{Planck Collaboration XXVIII}. 2014, \aap, 571, A28

\bibitem[{{Puglisi} \& {Bai}(2020)}]{PUG20}
{Puglisi}, G. \& {Bai}, X. 2020, \apj, 905, 143

\bibitem[{Puglisi {et~al.}(2018)Puglisi, Galluzzi, Bonavera, Gonzalez-Nuevo,
  Lapi, Massardi, Perrotta, Baccigalupi, Celotti, \& Danese}]{PUG18}
Puglisi, G., Galluzzi, V., Bonavera, L., {et~al.} 2018, The Astrophysical
  Journal, 858, 85

\bibitem[{{Remazeilles} {et~al.}(2018){Remazeilles}, {Banday}, {Baccigalupi},
  {Basak}, {Bonaldi}, {De Zotti}, {Delabrouille}, {Dickinson}, {Eriksen},
  {Errard}, {Fernandez-Cobos}, {Fuskeland}, {Herv{\'\i}as-Caimapo},
  {L{\'o}pez-Caniego}, {Martinez-Gonz{\'a}lez}, {Roman}, {Vielva}, {Wehus},
  {Achucarro}, {Ade}, {Allison}, {Ashdown}, {Ballardini}, {Banerji},
  {Bartlett}, {Bartolo}, {Baumann}, {Bersanelli}, {Bonato}, {Borrill},
  {Bouchet}, {Boulanger}, {Brinckmann}, {Bucher}, {Burigana}, {Buzzelli},
  {Cai}, {Calvo}, {Carvalho}, {Castellano}, {Challinor}, {Chluba}, {Clesse},
  {Colantoni}, {Coppolecchia}, {Crook}, {D'Alessandro}, {de Bernardis}, {de
  Gasperis}, {Diego}, {Di Valentino}, {Feeney}, {Ferraro}, {Finelli},
  {Forastieri}, {Galli}, {Genova-Santos}, {Gerbino}, {Gonz{\'a}lez-Nuevo},
  {Grandis}, {Greenslade}, {Hagstotz}, {Hanany}, {Handley},
  {Hernandez-Monteagudo}, {Hills}, {Hivon}, {Kiiveri}, {Kisner}, {Kitching},
  {Kunz}, {Kurki-Suonio}, {Lamagna}, {Lasenby}, {Lattanzi}, {Lesgourgues},
  {Lewis}, {Liguori}, {Lindholm}, {Luzzi}, {Maffei}, {Martins}, {Masi},
  {Matarrese}, {McCarthy}, {Melin}, {Melchiorri}, {Molinari}, {Monfardini},
  {Natoli}, {Negrello}, {Notari}, {Paiella}, {Paoletti}, {Patanchon}, {Piat},
  {Pisano}, {Polastri}, {Polenta}, {Pollo}, {Poulin}, {Quartin},
  {Rubino-Martin}, {Salvati}, {Tartari}, {Tomasi}, {Tramonte}, {Trappe},
  {Trombetti}, {Tucker}, {Valiviita}, {Van de Weijgaert}, {van Tent}, {Vennin},
  {Vittorio}, {Young}, \& {Zannoni}}]{REM18}
{Remazeilles}, M., {Banday}, A.~J., {Baccigalupi}, C., {et~al.} 2018, \jcap,
  2018, 023

\bibitem[{{Rubi{\~n}o-Mart{\'\i}n} {et~al.}(2012){Rubi{\~n}o-Mart{\'\i}n},
  {Rebolo}, {Aguiar}, {G{\'e}nova-Santos}, {G{\'o}mez-Re{\~n}asco}, {Herreros},
  {Hoyland}, {L{\'o}pez-Caraballo}, {Pelaez Santos}, {Sanchez de la Rosa},
  {Vega-Moreno}, {Viera-Curbelo}, {Mart{\'\i}nez-Gonzalez}, {Barreiro},
  {Casas}, {Diego}, {Fern{\'a}ndez-Cobos}, {Herranz}, {L{\'o}pez-Caniego},
  {Ortiz}, {Vielva}, {Artal}, {Aja}, {Cagigas}, {Cano}, {de la Fuente},
  {Mediavilla}, {Ter{\'a}n}, {Villa}, {Piccirillo}, {Battye}, {Blackhurst},
  {Brown}, {Davies}, {Davis}, {Dickinson}, {Harper}, {Maffei}, {McCulloch},
  {Melhuish}, {Pisano}, {Watson}, {Hobson}, {Grainge}, {Lasenby}, {Saunders},
  \& {Scott}}]{Rub12}
{Rubi{\~n}o-Mart{\'\i}n}, J.~A., {Rebolo}, R., {Aguiar}, M., {et~al.} 2012,
  Society of Photo-Optical Instrumentation Engineers (SPIE) Conference Series,
  Vol. 8444, {The QUIJOTE-CMB experiment: studying the polarisation of the
  galactic and cosmological microwave emissions}, 84442Y

\bibitem[{Rumelhart {et~al.}(1986)Rumelhart, Hinton, \& Williams}]{Rum86}
Rumelhart, D.~E., Hinton, G.~E., \& Williams, R.~J. 1986, Nature, 323, 533

\bibitem[{{Saikia} \& {Shastri}(1984)}]{SAI84}
{Saikia}, D.~J. \& {Shastri}, P. 1984, \mnras, 211, 47

\bibitem[{{Starobinski{\v{i}}}(1979)}]{STA79}
{Starobinski{\v{i}}}, A.~A. 1979, Soviet Journal of Experimental and
  Theoretical Physics Letters, 30, 682

\bibitem[{{Trombetti} {et~al.}(2018){Trombetti}, {Burigana}, {De Zotti},
  {Galluzzi}, \& {Massardi}}]{TRO18}
{Trombetti}, T., {Burigana}, C., {De Zotti}, G., {Galluzzi}, V., \& {Massardi},
  M. 2018, \aap, 618, A29

\bibitem[{{Tucci} \& {Toffolatti}(2012)}]{TUC12}
{Tucci}, M. \& {Toffolatti}, L. 2012, Advances in Astronomy, 2012, 624987

\bibitem[{{Tucci} {et~al.}(2011){Tucci}, {Toffolatti}, {de Zotti}, \&
  {Mart{\'\i}nez-Gonz{\'a}lez}}]{TUC11}
{Tucci}, M., {Toffolatti}, L., {de Zotti}, G., \& {Mart{\'\i}nez-Gonz{\'a}lez},
  E. 2011, \aap, 533, A57

\bibitem[{{W}es {M}c{K}inney(2010)}]{pandas}
{W}es {M}c{K}inney. 2010, in {P}roceedings of the 9th {P}ython in {S}cience
  {C}onference, ed. {S}t\'efan van~der {W}alt \& {J}arrod {M}illman, 56 -- 61

\bibitem[{Zonca {et~al.}(2019)Zonca, Singer, Lenz, Reinecke, Rosset, Hivon, \&
  Gorski}]{zon19}
Zonca, A., Singer, L.~P., Lenz, D., {et~al.} 2019, Journal of Open Source
  Software, 4, 1298

\end{thebibliography}


\end{document}